\documentclass[reprint, aps, amsmath, amssymb, superscriptaddress]{revtex4-2}

\usepackage{graphicx}
\usepackage{amsthm}
\usepackage{amsmath}
\usepackage{url}
\usepackage[left]{lineno}
\usepackage{color}
\usepackage{ulem}
\usepackage[symbol]{footmisc}

\begin{document}
\title{Differential cross sections and photon beam asymmetries of $\eta$ photoproduction on the proton at $E_\gamma=1.3-2.4$ GeV}
\author{T.~Hashimoto}
\affiliation{\small Research Center for Nuclear Physics, Osaka University, Ibaraki, Osaka 567-0047, Japan}
\affiliation{\small Department of Physics, Kyoto University, Kyoto 606-8502, Japan}

\author{T.~Nam}
\affiliation{\small Research Center for Nuclear Physics, Osaka University, Ibaraki, Osaka 567-0047, Japan}

\author{N.~Muramatsu}
\affiliation{\small Research Center for Electron Photon Science, Tohoku University, Sendai, Miyagi 982-0826, Japan}

\author{J.K.~Ahn}
\affiliation{\small Department of Physics, Korea University, Seoul 02841, Republic of Korea}

\author{W.C.~Chang}
\affiliation{\small Institute of Physics, Academia Sinica, Taipei 11529, Taiwan}

\author{J.Y.~Chen}
\affiliation{\small National Synchrotron Radiation Research Center, Hsinchu 30076, Taiwan}

\author{M.L.~Chu}
\affiliation{\small Institute of Physics, Academia Sinica, Taipei 11529, Taiwan}

\author{S.~Dat\'{e}}
\affiliation{\small Japan Synchrotron Radiation Research Institute (SPring-8), Sayo, Hyogo 679-5198, Japan}
\affiliation{\small Research Center for Nuclear Physics, Osaka University, Ibaraki, Osaka 567-0047, Japan}

\author{T.~Gogami}
\affiliation{\small Department of Physics, Kyoto University, Kyoto 606-8502, Japan}

\author{H.~Goto}
\affiliation{\small Research Center for Nuclear Physics, Osaka University, Ibaraki, Osaka 567-0047, Japan}

\author{H.~Hamano}
\affiliation{\small Research Center for Nuclear Physics, Osaka University, Ibaraki, Osaka 567-0047, Japan}

\author{Q.H.~He}
\affiliation{\small Department of Nuclear Science \& Engineering, College of Material  Science and Technology, Nanjing University of Aeronautics and Astronautics, Nanjing 210016, China}

\author{K.~Hicks}
\affiliation{\small Department of Physics and Astronomy, Ohio University, Athens, OH 45701, USA}

\author{T.~Hiraiwa}
\affiliation{\small RIKEN SPring-8 Center, Sayo, Hyogo 679-5148, Japan}

\author{Y.~Honda}
\affiliation{\small Research Center for Electron Photon Science, Tohoku University, Sendai, Miyagi 982-0826, Japan}

\author{T.~Hotta}
\affiliation{\small Research Center for Nuclear Physics, Osaka University, Ibaraki, Osaka 567-0047, Japan}

\author{H.~Ikuno}
\affiliation{\small Research Center for Nuclear Physics, Osaka University, Ibaraki, Osaka 567-0047, Japan}

\author{Y.~Inoue}
\affiliation{\small Research Center for Electron Photon Science, Tohoku University, Sendai, Miyagi 982-0826, Japan}

\author{T.~Ishikawa}
\affiliation{\small Research Center for Electron Photon Science, Tohoku University, Sendai, Miyagi 982-0826, Japan}

\author{I.~Jaegle}
\affiliation{\small Thomas Jefferson National Accelerator Facility, Newport News, Virginia 23606, USA}

\author{J.M.~Jo}
\affiliation{\small Department of Physics, Korea University, Seoul 02841, Republic of Korea}

\author{Y.~Kasamatsu}
\affiliation{\small Research Center for Nuclear Physics, Osaka University, Ibaraki, Osaka 567-0047, Japan}

\author{H.~Katsuragawa}
\affiliation{\small Research Center for Nuclear Physics, Osaka University, Ibaraki, Osaka 567-0047, Japan}

\author{S.~Kido}
\affiliation{\small Research Center for Electron Photon Science, Tohoku University, Sendai, Miyagi 982-0826, Japan}

\author{Y.~Kon}
\affiliation{\small Research Center for Nuclear Physics, Osaka University, Ibaraki, Osaka 567-0047, Japan}

\author{S.~Masumoto}
\affiliation{\small Department of Physics, University of Tokyo, Tokyo 113-0033, Japan}

\author{Y.~Matsumura}
\affiliation{\small Research Center for Electron Photon Science, Tohoku University, Sendai, Miyagi 982-0826, Japan}

\author{M.~Miyabe}
\affiliation{\small Research Center for Electron Photon Science, Tohoku University, Sendai, Miyagi 982-0826, Japan}

\author{K.~Mizutani}
\affiliation{\small Thomas Jefferson National Accelerator Facility, Newport News, Virginia 23606, USA}

\author{T.~Nakamura}
\affiliation{\small Department of Education, Gifu University, Gifu 501-1193, Japan}

\author{T.~Nakano}
\affiliation{\small Research Center for Nuclear Physics, Osaka University, Ibaraki, Osaka 567-0047, Japan}

\author{M.~Niiyama}
\affiliation{\small Department of Physics, Kyoto Sangyo University, Kyoto 603-8555, Japan}

\author{Y.~Nozawa}
\affiliation{\small Department of Radiology, The University of Tokyo Hospital, Tokyo 113-8655, Japan}

\author{Y.~Ohashi}
\affiliation{\small Japan Synchrotron Radiation Research Institute (SPring-8), Sayo, Hyogo 679-5198, Japan}
\affiliation{\small Research Center for Nuclear Physics, Osaka University, Ibaraki, Osaka 567-0047, Japan}

\author{H.~Ohnishi}
\affiliation{\small Research Center for Electron Photon Science, Tohoku University, Sendai, Miyagi 982-0826, Japan}

\author{T.~Ohta}
\affiliation{\small Department of Radiology, The University of Tokyo Hospital,  Tokyo 113-8655, Japan}

\author{K.~Ozawa}
\affiliation{\small Institute of Particle and Nuclear Studies, High Energy Accelerator Research Organization (KEK), Tsukuba, Ibaraki 305-0801, Japan}

\author{C.~Rangacharyulu}
\affiliation{\small Department of Physics and Engineering Physics, University of Saskatchewan, Saskatoon, Canada SK S7N 5E2}

\author{S.Y.~Ryu}
\affiliation{\small Research Center for Nuclear Physics, Osaka University, Ibaraki, Osaka 567-0047, Japan}

\author{Y.~Sada}
\affiliation{\small Research Center for Electron Photon Science, Tohoku University, Sendai, Miyagi 982-0826, Japan}

\author{T.~Shibukawa}
\affiliation{\small Department of Physics, University of Tokyo, Tokyo 113-0033, Japan}

\author{H.~Shimizu}
\affiliation{\small Research Center for Electron Photon Science, Tohoku University, Sendai, Miyagi 982-0826, Japan}
\affiliation{\small Research Center for Nuclear Physics, Osaka University, Ibaraki, Osaka 567-0047, Japan}

\author{R.~Shirai}
\affiliation{\small Research Center for Electron Photon Science, Tohoku University, Sendai, Miyagi 982-0826, Japan}

\author{K.~Shiraishi}
\affiliation{\small Research Center for Electron Photon Science, Tohoku University, Sendai, Miyagi 982-0826, Japan}

\author{E.A.~Strokovsky}
\affiliation{\small Laboratory of High Energy Physics, Joint Institute for Nuclear Research, Dubna, Moscow Region, 142281, Russia}
\affiliation{\small Research Center for Nuclear Physics, Osaka University, Ibaraki, Osaka 567-0047, Japan}

\author{Y.~Sugaya}
\affiliation{\small Research Center for Nuclear Physics, Osaka University, Ibaraki, Osaka 567-0047, Japan}

\author{M.~Sumihama}
\affiliation{\small Department of Education, Gifu University, Gifu 501-1193, Japan}
\affiliation{\small Research Center for Nuclear Physics, Osaka University, Ibaraki, Osaka 567-0047, Japan}

\author{S.~Suzuki}
\affiliation{\small Japan Synchrotron Radiation Research Institute (SPring-8), Sayo, Hyogo 679-5198, Japan}

\author{S.~Tanaka}
\affiliation{\small Research Center for Nuclear Physics, Osaka University, Ibaraki, Osaka 567-0047, Japan}

\author{Y.~Taniguchi}
\affiliation{\small Research Center for Electron Photon Science, Tohoku University, Sendai, Miyagi 982-0826, Japan}

\author{A.~Tokiyasu}
\affiliation{\small Research Center for Electron Photon Science, Tohoku University, Sendai, Miyagi 982-0826, Japan}

\author{N.~Tomida}
\affiliation{\small Research Center for Nuclear Physics, Osaka University, Ibaraki, Osaka 567-0047, Japan}

\author{Y.~Tsuchikawa}
\affiliation{\small J-PARC Center, Japan Atomic Energy Agency, Tokai, Ibaraki 319-1195, Japan}

\author{T.~Ueda}
\affiliation{\small Research Center for Electron Photon Science, Tohoku University, Sendai, Miyagi 982-0826, Japan}

\author{H.~Yamazaki}
\affiliation{\small Radiation Science Center, High Energy Accelerator Research Organization (KEK), Tokai, Ibaraki 319-1195, Japan}

\author{R.~Yamazaki}
\affiliation{\small Research Center for Electron Photon Science, Tohoku University, Sendai, Miyagi 982-0826, Japan}

\author{Y.~Yanai}
\affiliation{\small Research Center for Nuclear Physics, Osaka University, Ibaraki, Osaka 567-0047, Japan}

\author{T.~Yorita}
\affiliation{\small Research Center for Nuclear Physics, Osaka University, Ibaraki, Osaka 567-0047, Japan}

\author{C.~Yoshida}
\affiliation{\small Research Center for Electron Photon Science, Tohoku University, Sendai, Miyagi 982-0826, Japan}

\author{M.~Yosoi}
\affiliation{\small Research Center for Nuclear Physics, Osaka University, Ibaraki, Osaka 567-0047, Japan}

\collaboration{LEPS2/BGOegg collaboration}
\begin{abstract}
 We have carried out exclusive measurements for the photoproduction of an $\eta$ meson from a proton target
 with an egg-shaped calorimeter made of BGO crystals (BGOegg) and forward charged-particle detectors at the SPring-8 LEPS2 beamline.
 The differential cross sections and photon beam asymmetries of the $\gamma p \to \eta p$ reaction are measured
 in a center-of-mass energy ($W$) range of $1.82$--$2.32$ GeV and a polar angle range of  $-1.0 < \cos{\theta^{\eta}_{\mathrm{c.m.}}} < 0.6$.
 The reaction is identified by selecting a proton and two $\gamma$'s produced by an $\eta$-meson decay.
 The kinematic fit method was employed to select the reaction candidate with the confidence level larger than $1$\%.
 A bump structure at $W$ = $2.0$--$2.3$ GeV in the differential cross section is confirmed at extremely backward $\eta$ polar angles,
 where the existing data are inconsistent with each other.
 This bump structure is likely associated with high-spin resonances that couple with $s\bar{s}$ quarks.
 The results of the photon beam asymmetries in a wide $\eta$ polar angle range are new for the photon beam energies exceeding $2.1$ GeV.
 These results are not reproduced by the existing partial wave analyses.
 They provide an additional constraint to nucleon resonance studies at high energies.
\end{abstract}

\maketitle
\clearpage
\section{Introduction}
 Most of the information about baryon resonances was obtained through $\pi$N scattering and photoproduction experiments with the help of partial wave analyses (PWAs).
 The currently observed mass spectra in the center-of-mass energy ($W$) region of $1$--$3$ GeV is summarized in Tables 80.1 and 80.2 of the Ref.~\cite{Zyla:2020zbs}.
 So far, constituent quark models are successful in reproducing the mass spectrum to a certain extent, especially in the energy range below $1.8$ GeV \cite{cqm}.
 However, the predicted resonance masses are often not consistent with experimental results.
 For instance, the Roper resonance N(1440)$1/2^+$ is calculated to have a  mass larger than the first negative parity baryon N(1535)$1/2^-$ in the quark models \cite{RoperN1440}.
 In addition, at the energies above $W$ = $1.8$ GeV, the number of experimentally established resonances is smaller than that of predicted states.
 The mass spectra are sensitive to the hadron structure beyond the existing constituent quark models.
 Therefore, in order to have a better understanding of QCD at low energies, it is necessary to clarify the mass spectra from the experimental side.
\par

 In the $\pi$N scattering experiments, an energized pion beam excites a nucleon to a higher state,
 but the excited states are influenced by the flavors of constituent quarks in the pion.
 In contrast, the photoproduction experiments use a high-energy photon beam which can couple to a quark-antiquark pair of any flavor,
 and the photon is able to produce excited states whose coupling to the $\pi$N scattering is weak.
 In addition, a photon beam can be highly polarized, providing an advantage in obtaining the spin information of intermediate resonances.
 Photoproduction experiments are thus getting more popular in modern baryon studies.
 Here the photoproduction of an $\eta$ meson is a prime example of such research subjects.
 The $\eta$ meson photoproduction has a reasonably large cross section and it offers an attractive capability of coupling with the $s\bar{s}$ component in an \textit{s}-channel baryon resonance.
 Moreover, the $\eta$ meson is an isoscalar particle and therefore can only couple to isospin 1/2 resonances.
 The $\eta$ photoproduction works as an isospin filter for the complex baryon resonance spectrum.
\par

 Several experimental results of the $\eta$ photoproduction on the proton around $W$ = 2 GeV have been published in the last two decades.
 The LEPS collaboration reported differential cross sections of the $\eta$ photoproduction at backward polar angles in 2009 \cite{leps2009}.
 Measurements of differential cross sections in a wide angular region were also published by the CLAS \cite{clas2009} and CBELSA/TAPS \cite{cbelsa2009} collaborations in 2009.
 The CLAS and CBELSA/TAPS experiments also showed the result of the photon beam asymmetries in 2017 \cite{clas2017} and 2020 \cite{cbelsa2020}, respectively.
 The LEPS collaboration has claimed the existence of a bump structure in the differential cross section above $W$ = $2$ GeV at the backward $\eta$ angles.
 A similar structure was also seen in the CLAS and CBELSA/TAPS experiments.
 However, the shapes and strengths of the bump structure in their differential cross sections are significantly inconsistent
 with each other, causing a controversial situation.
 In particular, the LEPS and CBELSA/TAPS experiments show differential cross sections larger than the CLAS result in the bump region.
 The LEPS data were obtained by a missing mass analysis with the detection of only a forward proton in the final state of the reaction.
 The measured angular region was limited to $\cos{\theta^{\eta}_{\mathrm{c.m.}}}<-0.6$.
 Therefore a systematic comparison with other experiments in the wide angular region is difficult.
 The CLAS experiment has achieved high statistical precision and wide angular coverage with the detection of both an $\eta$ meson and a proton in the final state.
 However, the acceptance at the extremely backward angles, which are the main focus for the study of the bump structure, is limited to above $\cos{\theta^{\eta}_{\mathrm{c.m.}}}=-0.855$.
 The CBELSA/TAPS experiment covers a wide angular region including the missing backward $\eta$ angles of the CLAS experiment.
 Nevertheless, their differential cross sections have relatively larger statistical uncertainties than the other experimental results.
 In addition, the differential cross sections of the CBELSA/TAPS experiment are systematically larger than those of the CLAS measurement in wide energy and angular bins.
\par

 In this article, we report our results on the differential cross sections and photon beam asymmetries
 of the $\eta$ photoproduction in a wide polar angle region $-1<\cos{\theta^{\eta}_{\mathrm{c.m.}}} < 0.6$.
 Both a proton and an $\eta$ meson decaying into two $\gamma$'s are detected
 so that we can examine the nature of the bump structure with the clear identification of $\eta$-photoproduction signals.
 At the most backward angles, the data were collected with high statistics to achieve more reliable studies that could not be ever done by the previous experiments.
 The beam asymmetry data for the total energies above $2.1$ GeV is presented for the first time
 thanks to the highly polarized photon beam, whose linear polarization is more than $90$\% above $W = 2.1$ GeV.
 This new data will provide crucial information for partial wave analyses including the bump study.
 Several neutral-meson photoproduction reactions can be measured simultaneously by using the large-acceptance calorimeter, called BGOegg,
 for a comprehensive comparison of baryon resonance contributions among different reaction modes.
 The BGOegg calorimeter currently has the world's highest resolution, enabling the acquisition of high-quality data.
\par

 This paper is organized as follows;
 In Section II, we describe our experimental setup.
 The data-analysis procedures are provided in Section III.
 In Section IV, the method to obtain the differential cross sections and photon beam asymmetries are presented in detail.
 The experimental results are shown in Section V.
 In Section VI, we discuss the obtained results and the comparison of them with several PWA calculations.
 Section VII is the summary of the present measurement and results.

\section{Experimental setup}
\begin{figure*}[!tp]
 \centering
 \includegraphics[width=172truemm]{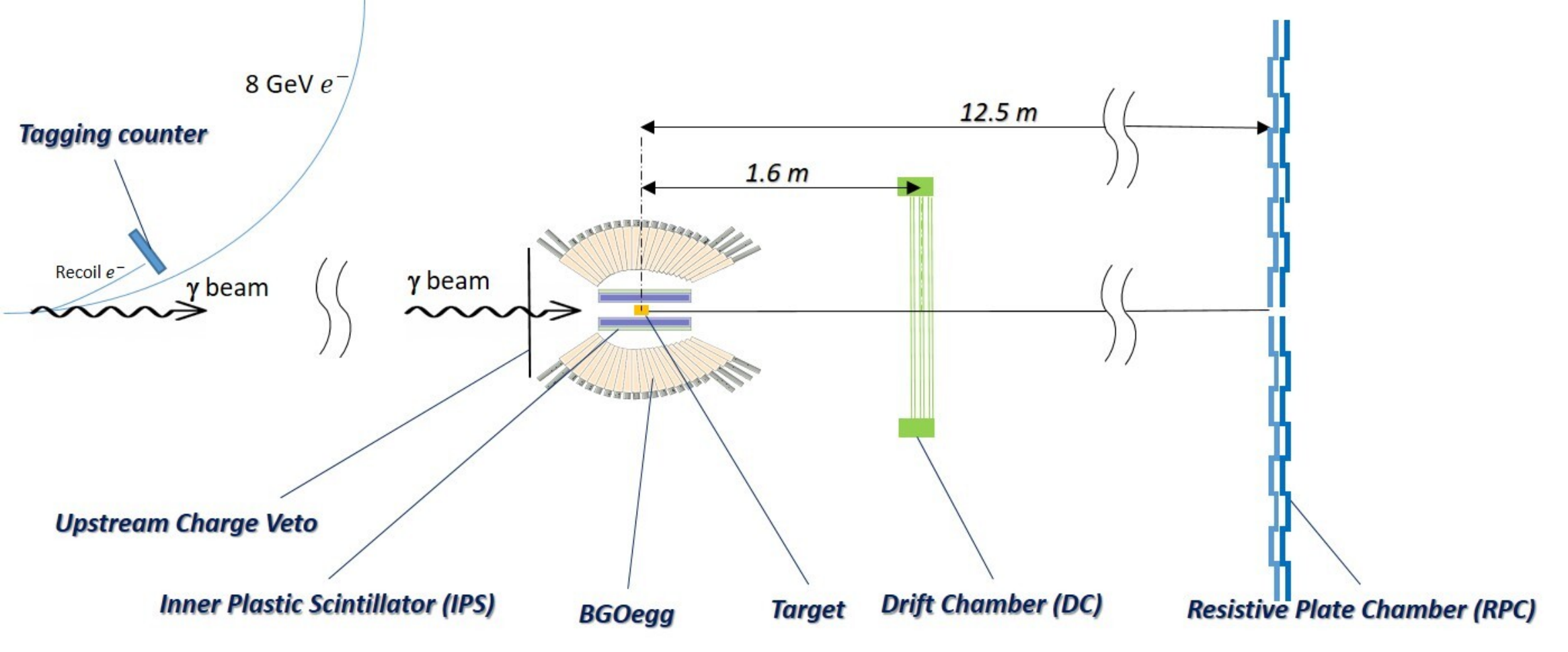}
 \caption{A schematic drawing of BGOegg experimental setup at the LEPS2 beamline (Top view).}
 \label{fig:exp_schem}
\end{figure*}

 We  carried out an experiment to study the $\eta$ photoproduction (the BGOegg experiment) using the LEPS2 beamline at SPring-8.
 Details of the LEPS2 beamline are described in Ref.~\cite{leps2}.
 Figure \ref{fig:exp_schem} shows the schematic view of detectors in the BGOegg experiment.

 A high-energy photon beam was produced by the backward Compton scattering of $355$ nm wavelength ultraviolet laser light from an $8$-GeV electron in the storage ring \cite{beamline}.
 Four laser beams can be injected simultaneously from the oscillators whose maximum output power was either of $16$ or $24$ W.
 The maximum energy of the scattered photon is 2.39 GeV at the Compton edge.
\par

 The energy of a backwardly scattered photon is determined by the tagging detector system (tagger), which is located downstream of a bending magnet of the storage ring.
 The tagger consists of two layers of $1$ mm-wide scintillating fiber bundles and two layers of $8$ mm-wide plastic scintillators
 to reconstruct the track of a recoil electron from the Compton scattering.
 Because the recoil electron has lost a part of the energy, its flight trajectory in the magnetic field deviates from that of the $8$-GeV electron orbit.
 The recoil electron momentum can be determined by analyzing its hit position on the tagger.
 The photon energy is then calculated event by event using the 4-momentum conservation law.
 The trigger signal for data acquisition is formed from a coincidence signal produced by
 two 8-mm wide plastic scintillators in a pair together with a requirement that at least two BGOegg crystals have hits.
 The hit rate of the tagger logic signal is counted by a scaler to monitor the intensity of a produced photon beam.
 The beam intensity was in the range of $1$--$1.8$ $\times 10^{6}$ photon/\textit{s} during the experiment for the present data set.
\par

 A scintillating counter with an effective area of $620 \times 620$ $\mathrm{mm^{2}}$ and a thickness of $3$ mm was installed just upstream of the BGOegg calorimeter.
 It was used to veto the $e^+e^-$ pair contaminating the photon beam.
 The target cell, made of thin polyimide films in the form of a cylindrical shape, is  placed in the center of the BGOegg calorimeter.
 A refrigerator which is connected with a hydrogen gas tank liquefies a part of the sealed gas,
 and fills the target cell with liquid hydrogen.
 The measured thickness of the target cell was $54$ mm, and the center of the target was shifted $3$ mm upstream
 from the designed center position due to an expansion effect of the target cell.
 This small shift affects the polar angle measurement of final reaction products and is taken into account in the offline analysis.
\par

 The BGOegg calorimeter consists of $1320$ BGO crystals with 20 radiation lengths, covering polar angles from $24$ to $144$ degrees.
 The crystals are distributed into $22$ layers in the polar angle direction with a ring of 60 crystals each.
 No frame material is inserted between the crystals.
 The energy calibration for each crystal was done by iteration so that the invariant mass of two $\gamma$'s,
 one of which deposits the largest fraction of its energy to the calibrated crystal, has a peak at the nominal $\pi^{0}$ mass \cite{Zyla:2020zbs}.
 The energy resolution of the BGOegg calorimeter has been evaluated to be $1.4$\% at the incident $\gamma$ energy of $1$ GeV \cite{bgoegg_nim}.
 The invariant mass resolution of the $\pi^0$ is $6.7$ MeV/$c^{2}$ with a $20$ mm thick carbon target.
 These resolutions are the world's best among the experiments conducted in a similar energy range.
 The identification of a charged or $\gamma$ signal for the BGOegg calorimeter hit was performed using the inner plastic scintillator (IPS).
 The IPS is composed of $30$ scintillator slabs which are $453$ mm long and $5$ mm thick.
 These slabs are arranged in a cylindrical shape around the target.
\par

\begin{figure*}[!tp]
  \centering
  \includegraphics[width=86truemm]{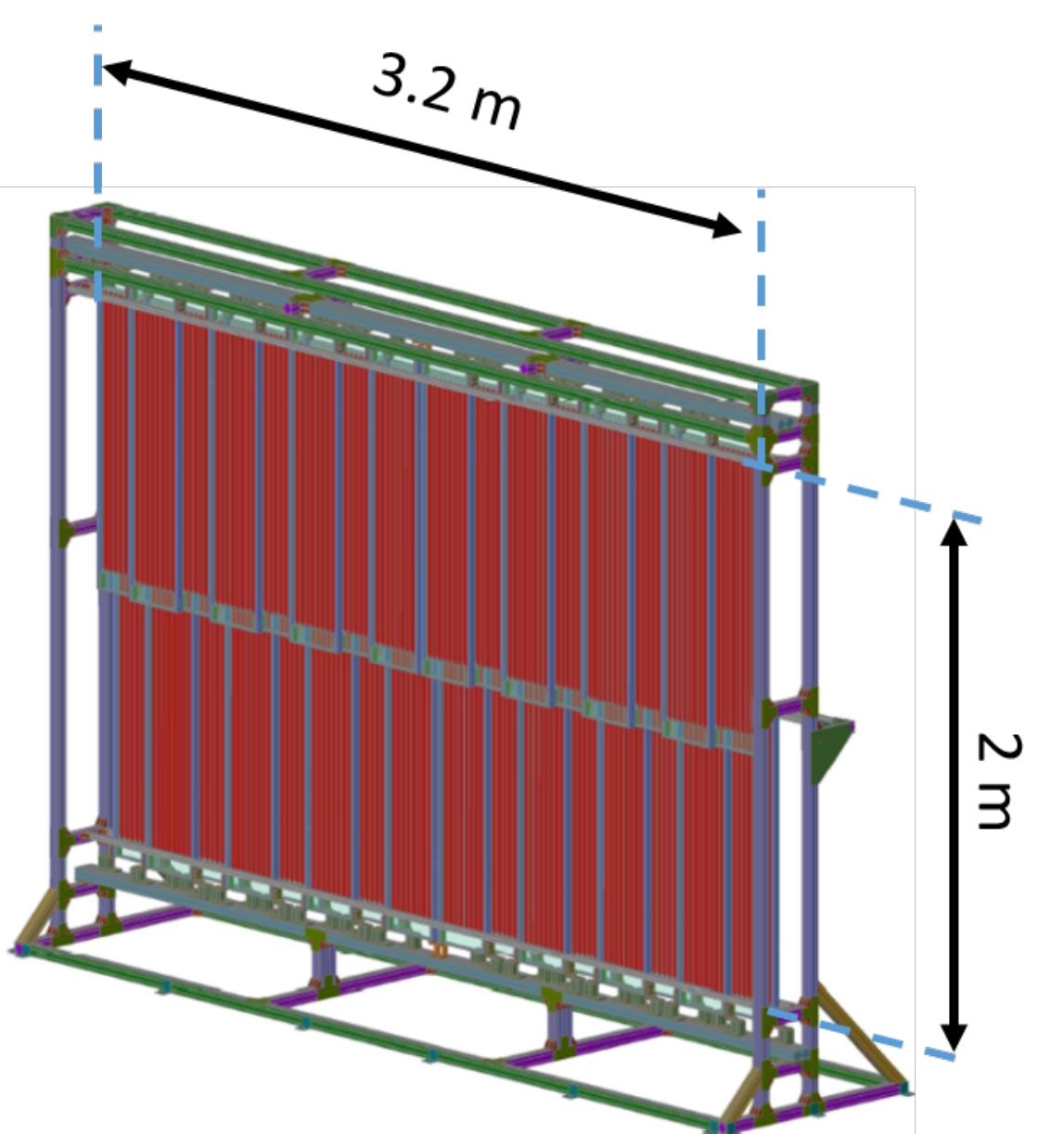}
  \caption{A 3-dimensional figure of ToF wall that is formed by $32$ RPCs.}
  \label{fig:rpc}
\end{figure*}

 Charged particles that were emitted to the forward open hole of the BGOegg calorimeter were detected using a Drift Chamber (DC).
 The DC consists of six separated planes.
 Each plane has 80 sense-wires with a wire interval of $16$ mm.
 These six planes are divided into three groups by the directions of sense wires,
 which is tilted at an azimuthal angle of 60 degrees relative to the other groups.
 The positions of sense wires in a certain plane are shifted by 8 mm relative to those in the other plane belonging to the same group.
 The position resolution of a DC hit on each plane is about $300$ $\mathrm{\mu m}$.
 The DC is located $1.6$ m downstream of the target, covering polar angles less than $21$ degrees.
\par

 At the $12.5$ m distance from the target, there is a Time-of-Flight (ToF) wall to measure the momentum of a proton, which is emitted at extremely forward angles.
 As shown in Fig.~\ref{fig:rpc}, the ToF wall consists of $32$ Resistive Plate Chambers (RPCs), each of which is $250$-mm wide and $1000$-mm long \cite{rpc_jinst, rpc_jinst2}.
 There are eight readout strips along the vertical direction in a chamber,
 and hit signals are read at both top- and bottom-ends.
 Figure \ref{fig:rpc} shows a size of the ToF wall and the arrangement of RPCs.
 The RPCs cover the laboratory polar angles less than $6.8$ degrees,
 which correspond to the most backward $\eta$ polar angles in the center-of-mass frame of the reaction, $\cos \theta^{\eta}_{\mathrm{c.m.}} < -0.95$.
 The time resolution of the RPC is $60$--$90$ ps providing a good momentum resolution less than $1$\% for an incident proton of $2$ GeV/c.
 The RPC allows the measurement of the differential cross sections of $\eta$ photoproduction at the most backward angles
 with full kinematic information, which makes our analysis more reliable.
 This extreme angular region is either inaccessible or associated with large uncertainties in other experiments.

\section{Data analysis}
\subsection{Event reconstruction}
 The production of a beam photon by a Compton scattering is identified offline
 if a recoil electron track is successfully reconstructed at the tagger with strict geometrical conditions.
 At first, the recoil electron must hit one or two layers of the scintillating fibers and two layers of the plastic scintillators,
 following one of the hit patterns that are pre-defined as possible geometrical arrangement for a straight track.
 Secondly, tight cuts are applied for the timing difference among plastic scintillators and scintillating fibers.
 The width of each fiber is $1$ mm but the second fiber layer is horizontally shifted by $0.5$ mm from the first layer
 to achieve a fine detector resolution.
 For the measurement of differential cross sections,
 the tagger reconstruction efficiency was evaluated to compensate for the signal loss
 due to track reconstruction failure in the offline analysis, multi-track detection in the tagger, and inefficiencies of tagger fibers.
 The reconstruction efficiency varies from $0.86$ to $0.93$ depending on the photon beam energy.
 The typical uncertainty of this reconstruction efficiency is $0.7$\%.
 A part of recoil electrons with high momenta hit a wall structure inside the vacuum chamber upstream of the tagger or the radiation shield box containing the tagger.
 Such a hit generates an electromagnetic shower, which produces a fake signal at the tagger.
 The shower contamination rate in the tagger triggers was estimated to be $0.0424 \pm 0.0006$ \cite{eggpi0}.
 This contamination was sufficiently removed offline by the tight geometrical conditions in the tagger reconstruction.
\par

 The photon beam energy was measured event-by-event from the hit position of a recoil electron at the tagger scintillating fibers.
 The photon beam energy was calibrated by using the kinematic information that was obtained from the detectors other than the tagger.
 In the offline analysis, it is possible to predict the photon beam energy from the kinematic fit of the reaction $\gamma p \to \pi^{0} \pi^{0} p$ without the tagger information.
 A polynomial function was fitted to the predicted energies depending on the tagger hit position in the calibration.
 Simultaneously, the photon beam energy resolution was estimated to be $12.1$ MeV.
 This resolution is predominantly influenced by the electron beam emittance.
 In addition, the consistency between the measured and predicted photon energies was examined to evaluate the tagger reconstruction efficiency mentioned above.
\par

 The measurement of differential cross sections needs an accurate determination of the photon beam flux.
 The photon beam flux was derived from the number of hits at the tagger plastic scintillators (tagger trigger logic signals).
 The tagger trigger counting is influenced by dead time because of a finite signal width (20 ns).
 The dead time depends on the tagger trigger rate and the electron filling pattern at SPring-8 \cite{SP8filling}.
 The typical dead time was $10$\%.
 The integrated counts of tagger triggers used for the present analysis reach $3.593 \times 10^{12}$ after correcting the dead time.
\par

 Due primarily to the effects of pair creations at materials in the long beamline from the Compton scattering point to the target,
 the number of beam photons counted by the tagger is not equal to the number of photons reaching the target.
 This loss must be corrected by determining the transmission rate of the photon beam.
 Based on the amount of materials on the beamline, this rate was estimated to be $0.772$.
 Then, the ratio of $\pi^{0}$ yields to tagger trigger counts was monitored to examine the run dependence of the transmission rate.
 It turned out that this ratio was changed when we tuned the laser injection causing a shift of the Compton scattering point.
 The amount of change was greater at the lower energies, where the Compton scattering produces photons with larger cone angles.
 It means that a part of scattered photons cannot go through the collimator on the beamline
 if the laser focal length to the Compton scattering point becomes longer than the designed distance.
 Therefore, the transmission rate was further corrected by multiplying an additional factor $F_{\mathrm{trans}}$, which depends on the photon beam energy $E_{\gamma}$ in GeV;
\begin{align}
 \label{eq:trans}
 F_{\mathrm{trans}}(E_{\gamma}) = 1 &+ 1.206 \times 10^{-2} \times (2.30 - E_{\gamma}) \notag \\
                                    &- 0.1113 \times (2.30 - E_{\gamma})^{2}
\end{align}
 This correction factor was evaluated by fitting a second-order polynomial function to the energy-dependent ratio of inclusive $\pi^{0}$ yields to tagger photon counts
 after this ratio was normalized in individual energy bins based on the value during the period with a good Compton scattering point.
 The beam loss was not observed at higher energies so that this correction factor was renormalized to 1 at the highest energy region.
\par

 Two $\gamma$'s from an $\eta$ photoproduction reaction are detected using the BGOegg calorimeter.
 The Moliere radius for BGO is 22.3 mm, which is
 a little larger than the front size of individual BGO crystals.
 Therefore, an electromagnetic shower of a $\gamma$ leaves its energy in multiple crystals around a core where the $\gamma$ is incident.
 The crystals with energy deposits are grouped into a ``cluster''.
 This cluster consists of several main crystals whose energies are greater than the discriminator threshold at about 10 MeV
 and neighboring peripheral crystals with smaller energies.
 The cluster energy was a sum of all the cluster members.
 The crystal with the largest energy was adopted as the core crystal of the cluster.
 The cluster timing was determined by using the core crystal.
 The center of a cluster was evaluated from the energy-weighted average of hit crystal positions.
 A charge of the cluster was identified by examining an IPS hit on the line between the target and the BGOegg cluster center.
 Thus, a proton from the $\eta$ photoproduction is also detectable as a charged particle at the BGOegg calorimeter.
 The detection efficiency for a proton of the IPS was estimated to be $0.9863 \pm 0.0009$.
\par

 The DC measures only the direction of a charged particle under no magnetic field.
 A straight line was fitted to each track candidate which contains five or more layer hits.
 The fit was performed by taking into account drift distances from individual hit wires and using an additional constraint by the target position.
 The tracks with $\chi^2$ probability greater than $1$\% were accepted for further analysis.
 The efficiency of finding a DC track, including both detection and reconstruction efficiencies, was estimated to be $0.9824 \pm 0.0044$
 by analyzing photoproduction reactions with a forward proton independently detected using the RPC.
\par

\begin{figure*}[!tp]
 \centering
 \includegraphics[width=172truemm]{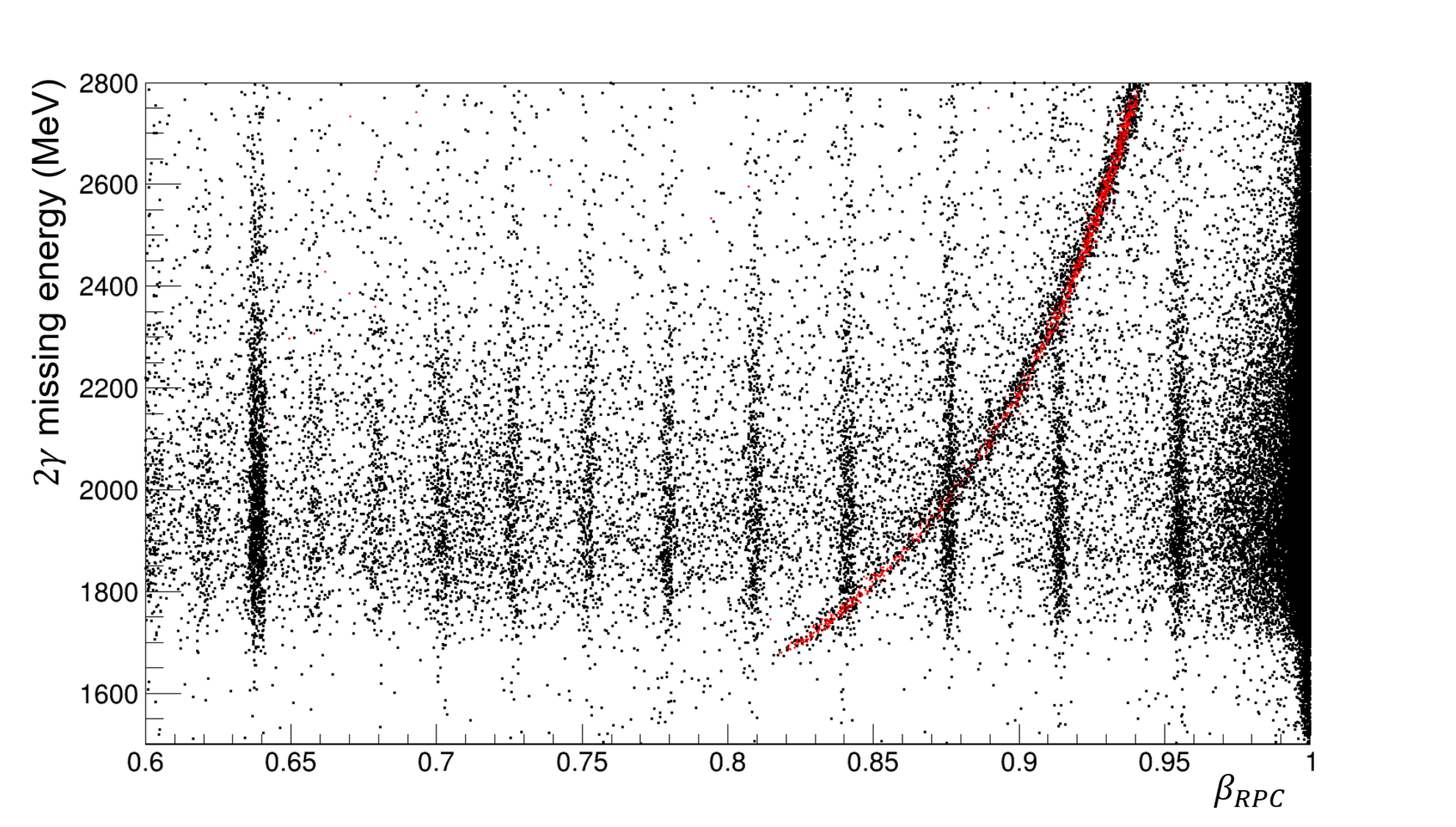}
 \caption{
 The correlation between the missing energy of a $\gamma\gamma$ pair detected at the BGOegg calorimeter and the velocity of a charged particle measured using the RPC.
 The black dots show the data before the kinematic fit
 but after requiring the existence of two neutral clusters at the BGOegg calorimeter with a loose cut on the missing mass.
 The vertical bands come from electron events, which are originated from the different electron bunches in the SPring-8 storage ring.
 The red dots are the data after passing the $99$\% confidence level cut in the kinematic fit.
 }
 \label{fig:missEvsbeta}
\end{figure*}

 A proton hit at the RPC wall is searched for around the position that is on the extension line of the reconstructed DC track.
 We obtained the hit position in the horizontal direction from the position of the hit strip.
 We evaluated the hit position in the vertical direction from the timing difference of signals from the top- and bottom-ends of a strip.
 The position resolutions in the horizontal and vertical directions were $7.5$ and $16$ mm, respectively.
 The reconstruction efficiency of an RPC hit was $0.931\pm0.023$.
 The velocity of a charged particle was measured from the time-of-flight (ToF) and
 the corresponding momentum was determined by assuming the proton mass for the detected particle.
 Protons are well separated from charged pions and electrons even by the velocity information if it is combined with the measurement at the BGOegg calorimeter.
 Figure \ref{fig:missEvsbeta} shows the correlation between the missing energy of a $\gamma\gamma$ pair detected using the BGOegg calorimeter
 and the velocity of a charged particle measured using the RPC.
 While events at the reconstruction level are shown by black dots, only the proton band (red dots) remains after the kinematic fit selection
 of $\gamma p \to \eta p \to \gamma\gamma p$ events with the use of both the RPC and the BGOegg calorimeter.

\subsection{Event selection}
 In the present analysis, the $\gamma p \rightarrow \eta p$ events were extracted from the data using the liquid hydrogen target.
 The $\eta$ meson was detected using the BGOegg calorimeter in the decay mode into $\gamma\gamma$, whose branching fraction is $0.3941 \pm 0.0020$ \cite{Zyla:2020zbs}.
 Event selection conditions are basically the same as those in the published article on $\pi^{0}$ photoproduction \cite{eggpi0}.
\par

 Events that have two neutral clusters at the BGOegg calorimeter were selected as signal candidates,
 if the difference between individual cluster and trigger timings was less than 3 ns.
 Neutral clusters whose central crystal was found at the most forward or backward edge layer of the BGOegg calorimeter were unused
 because we were not able to measure the correct cluster energy due to a leak.
 The minimum energy of each cluster was required to be 50 MeV in order to remove the accidental hits.
\par

 In addition to the neutral clusters, a proton was detected in the wider acceptance defined by a combination of
 the BGOegg calorimeter, the DC, and the RPC depending on the emission angle.
 Proton emission angles in the range of $24 < \theta^{p}_{\textrm{lab}} < 144 \textrm{ degrees } (-0.5 < \cos{\theta^{\eta}_{\mathrm{c.m.}}} < 0.6)$ was covered by the BGOegg calorimeter,
 where the direction was measured based on a line from the target center to the charged cluster core.
 The timing and minimum energy conditions of charged clusters were the same as those of neutral clusters.
 Unlike the neutral cluster, the charged clusters whose core was found at the edge layers of the BGOegg calorimeter were used
 because we used only the emission angle information.
 Protons emitted at the angles $\theta^{p}_{\textrm{lab}} < 21 \textrm{ degrees } (\cos{\theta^{\eta}_{\mathrm{c.m.}}} < -0.5)$ were measured using the DC.
 In the case of extremely forward angles $\theta^{p}_{\textrm{lab}} < 6.8 \textrm{ degrees } (\cos{\theta^{\eta}_{\mathrm{c.m.}}} < 0.95)$,
 we performed an additional analysis by using the events where protons were detected using both the DC and the RPC.
 The total number of charged tracks in a reconstructed event was limited to 1.
\par

\begin{figure}[!tbp]
 \centering
 \includegraphics[width=86truemm]{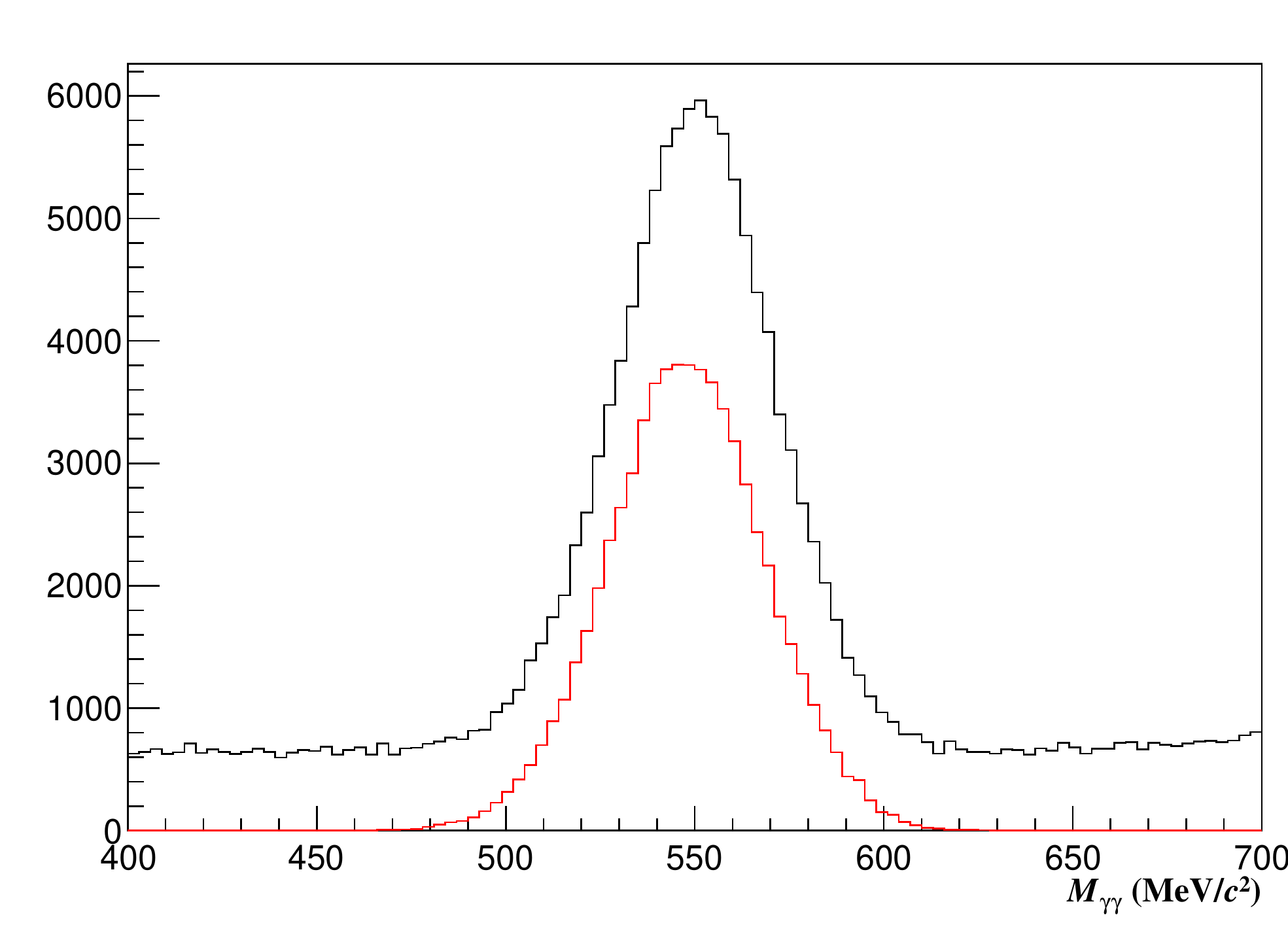}
 \caption{The invariant mass distributions for $\gamma\gamma$ pairs detected at the BGOegg calorimeter.
          The black histogram shows the events surviving after applying the loose condition that
          the missing mass of a $\gamma\gamma$ pair is less than 1200 $\mathrm{MeV}/c^{2}$.
          The red histogram shows the events that survive after applying the $99$\% confidence level cut.}
 \label{fig:inv_mass}
\end{figure}

 After measuring all the final state particles, a kinematic fit was performed by assuming the reaction $\gamma p \rightarrow \eta p \rightarrow \gamma\gamma p$.
 Five constraints are defined by a series of equations describing the 4-momentum conservation between the initial and final states of the reaction
 and the equality between the $\gamma\gamma$ invariant mass and the nominal $\eta$ mass.
 The measured energy, polar and azimuthal angles were varied within the uncertainties determined by detector resolutions.
 The vertex position was also floated with the constraint of the target size.
 The energy and angular resolutions of $\gamma$'s detected at the BGOegg calorimeter were estimated by generating single-photon events
 over the full ranges of energies and angles in the GEANT4 \cite{geant4} based simulation package.
 The generated $\gamma$'s were simulated with the detector setup implemented realistically.
 The relevant resolutions were obtained from the comparison of the true and simulated values.
 Figure \ref{fig:inv_mass} shows the invariant mass distributions for $\gamma\gamma$ pairs detected at the BGOegg calorimeter.
 After applying the $99$\% confidence level cut at the kinematic fit, we have succeeded in removing most of the background without acceptance loss.
 However, the background rejection was not complete because the proton momentum was not measured in the case of using the BGOegg calorimeter or the DC only.
 The background contamination is discussed in the next subsection.

\subsection{Yield extraction with background subtraction}
 In order to extract signal yields, we need to estimate background contributions in the event sample that remains after the selection described in the previous subsection.
 After the kinematic fit with a $99$\% confidence level cut, kinematical distributions of signals and backgrounds become too similar to be distinguished from each other.
 Therefore we examined amounts of individual background processes by performing a template fitting for the background-enhanced sample with loose event selection.
 The loose selection includes the cuts on the invariant and missing masses of a $\gamma\gamma$ pair to select an $\eta$ meson and a proton, respectively.
 An angular consistency between the detected proton and the missing momentum of a $\gamma\gamma$ pair was also required.
\par

\begin{figure*}[!tp]
 \centering
 \includegraphics[width=100truemm]{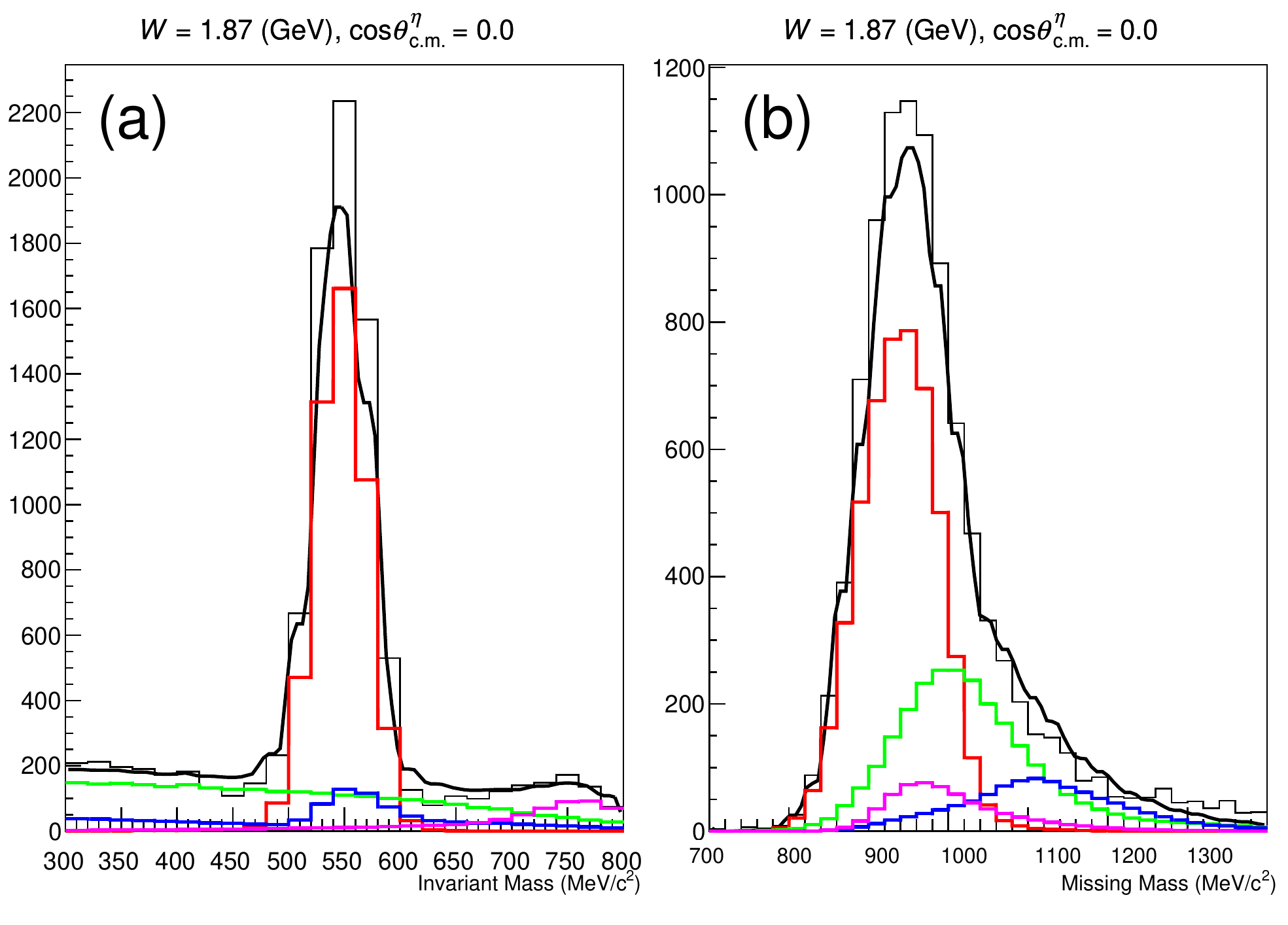}
 \caption{An example of the template fitting at the kinematic bin of $W$ = $1.87$ GeV and $\cos{\theta^{\eta}_{\mathrm{c.m.}}} = 0$.
          The black histogram in the panel (a) shows the invariant mass distribution for $\gamma\gamma$ pairs detected at the BGOegg calorimeter in the real data.
          The red, green, blue, and magenta histograms show the template mass spectra obtained from
          MC simulations of the $\eta, \pi^{0}\pi^{0}, \eta\pi^{0}, \textrm{and } \omega$ photoproduction processes, respectively.
          The thick black line shows a sum of all the template spectra.
          The black histogram in the panel (b) shows the missing mass distribution of a $\gamma\gamma$ pair in the real data.
          The line colors of fit results are defined in the same way as those in (a).
         }
 \label{fig:template_fit}
\end{figure*}

 We considered three background reactions in the template fitting: $\gamma p \to \pi^{0}\pi^{0} p$, $\gamma p \to \pi^{0}\eta p$, and $\gamma p \to \omega p$.
 The mesons in these reactions decay into multiple $\gamma$'s, and only two of the final-state $\gamma$'s are detected at the BGOegg calorimeter.
 At first, we prepared template histograms for the signal ($\gamma p \to \eta p$) and above background processes
 by using Monte Carlo (MC) simulations and applying the loose event selection that was also applied to the real data sample.
 Characteristic shapes for individual background processes appear in the side-band regions of the invariant and missing mass spectra for the detected $\gamma\gamma$ pairs.
 Therefore, the template histograms were prepared for these mass spectra and simultaneously fitted to the corresponding distributions in the real data.
 The fitting of template histograms was performed at individual kinematic bins, separated in total energies and polar angles for the measurement of differential cross sections and photon beam asymmetries.
\par

 Figure \ref{fig:template_fit} shows an example of the template fit in a certain kinematic bin.
 The colored histograms show the mass distributions from the $\eta, \pi^{0}\pi^{0}, \eta\pi^{0}, \textrm{and } \omega$ photoproduction, as described in the figure caption.
 The black line shows a sum of all the template spectra obtained by the fit.
 From this fitting result, the normalization factors of simulated background samples to the real data size were determined.
 The final contamination rate after the kinematic fit with the $99$\% confidence level cut
 was then evaluated by applying this condition to individual simulated background samples and taking into account the obtained normalization factors.
 The signal yields in individual kinematic bins were obtained by subtracting the estimated amount of backgrounds.

\par

 The number of events that survive after applying the $99$\% confidence level cut at the kinematic fit is $6.2 \times 10^{4}$ events.
 The background ratio is $3.1$--$36.9$\% depending on the kinematic bins.
 In the lower energy region, the background ratio at backward $\eta$ angles is larger than that at middle angles.
 On the other hand, the dependence is opposite in the higher energy region.
 The background ratio becomes larger at higher energies.
 After the background subtraction, the number of signal yields is estimated to be $5.5 \times 10^{4}$ events.

\subsection{Geometrical acceptance}
 The geometrical acceptance was obtained by the GEANT4 based MC simulation package developed for the BGOegg experiment.
 The $\gamma p \to \eta p$ events were generated with an isotropic angular distribution.
 The same event selection conditions as those in the real data analysis were applied to the generated sample for the acceptance measurement.
 The cross sections obtained from this acceptance were then fed back to the MC simulation so as to reflect the realistic kinematic distributions in a new round of acceptance calculation.
 This iteration process ended when the obtained differential cross sections became stable within $1$\% of the previous iteration values.
 The typical acceptance is $50$\% at backward $\eta$ angles and reduced at forward angles.
 There is detection sensitivity up to $\cos{\theta^{\eta}_{\mathrm{c.m.}}} \sim 0.6$.

\section{Measurement of differential cross sections and photon beam asymmetries}
\subsection{Differential cross section}
 The differential cross section $d\sigma / d\Omega$ was calculated using the following equation:
\begin{equation}
 \label{eq:dcs}
 \frac{d\sigma}{d\Omega} = \frac{Y_{\eta}}{N_{\gamma} \cdot T_{\gamma} \cdot F_{\mathrm{trans}} \cdot \rho_{N} \cdot A \cdot \mathrm{Br}_{\eta} \cdot \epsilon}\frac{1}{\Delta\Omega}
\end{equation}
 The differential cross sections were measured in 20 energy bins at $W$ = $1.82$--$2.32$ GeV and 16 polar angle bins at $\cos{\theta^{\eta}_{\mathrm{c.m.}}}$ = $-1.0$--$0.60$.
 $Y_{\eta}$ is the $\eta$ photoproduction yield in a certain kinematic bin, used for the cross section measurement.
 This value was obtained by counting the number of events after requiring the signal selection conditions and subtracting backgrounds as described in Sec.~III-B and C.
 $N_{\gamma}$ is the number of beam photons after the correction by the dead time, described in Sec.~III-A.
 $T_\gamma$ and $F_{\mathrm{trans}}$ are the transmission rate and the energy-dependent correction factor for it, respectively, as described in Sec.~III-A.
 $\rho_{N}$ is the number density of protons in the liquid hydrogen target (0.0708 $\textrm{g/cm}^3$).
 $A$ is the geometrical acceptance of the detector system for each energy and angular bin, described in Sec.~III-D.
 $\mathrm{Br}_{\eta}$ is the branching fraction of the $\eta \rightarrow \gamma\gamma$ ($0.3941$).
 $\epsilon$ is the product of other efficiency factors, namely,
 the tagger reconstruction efficiency, the fraction of true tagger tracks after removing shower contributions,
 and the proton detection efficiency at the IPS or the DC, all of which are described in Sec.~III-A.
\par

\begin{table}[!bp]
 \footnotesize
 \centering
 \caption{Systematic uncertainties of the differential cross section measurement}
 \begin{tabular}{l@{\hspace{10truemm}}l}\hline \hline
  Source of systematic uncertainty & Typical value \\ \hline
  Energy dependent transmission & \\
  ~~Fit function dependence & $0.2$--$1.0$\% \\
  ~~Normalization method & $2.8$\% \\
  ~~Energy dependence & $0.3$--$2.0$\% \\
  Target length & $1.3$\% \\
  Beam position shift & $0.01$--$8.8$\% \\
  Kinematic-fit cut dependence & $0.01$--$3.4$\% \\
  Tagger reconstruction efficiency & $0.57$--$0.92$\% \\
  Shower contribution & $1.4$\% \\
  Proton detection efficiency & $0.09$\% (IPS), $0.45$\% (DC) \\
  Branching ratio $(\eta \rightarrow \gamma \gamma)$ & $0.50$\% \\ \hline \hline
 \end{tabular}
 \label{table:dcs_sys}
\end{table}

 Systematic uncertainties for the measurement of differential cross sections are listed in Table \ref{table:dcs_sys}.
 These were estimated in the same way as the published analysis for the $\pi^{0}$ photoproduction \cite{eggpi0}.
 The ambiguities for the energy-dependent transmission and the target length are completely the same as those described in Ref.~\cite{eggpi0}.
 An influence of the transverse shift of the photon beam is also reported in Ref.~\cite{eggpi0}.
 The amount of the shift should be consistent with that in Ref.~\cite{eggpi0},
 but the effect on the geometrical acceptance depends on the angular distribution of each reaction.
 So we recalculated possible changes of the geometrical acceptance factors in individual kinematic bins by the MC simulation.
 The estimated variations of the cross section values were in the range of $0.01$--$8.8$\% depending on the kinematic bin.
 In the present analysis, we applied the $\chi^2$ probability cut at $1$\% to select signals after the kinematic fit.
 For estimating the uncertainty due to the cut point, the differential cross section was recalculated
 by requiring the $\chi^2$ probability above $5$\%, where the probability distribution is flat.
 The resulting changes were in the range of $0.01$--$3.4$\%.
 Uncertainties in measurement of the tagger reconstruction efficiency, the shower contribution, and the proton detection efficiency,
 described in Sec.~III-A, were reflected to the ambiguities of differential cross sections in the individual kinematic bins.
 The uncertainty of the branching fraction of the $\eta \to \gamma\gamma$ decay was also taken into account based on the Particle Data Group value \cite{Zyla:2020zbs}.
 The total systematic uncertainties were evaluated to be $3.3$--$11$\% by taking a root of the quadratic sum of the listed uncertainties.

\subsection{Photon beam asymmetry}
 In the pseudoscalar-meson photoproduction with a linearly polarized beam,
 the differential cross section has asymmetry depending on the azimuthal angle of the produced meson relative to the beam polarization direction.
 This is called photon beam asymmetry $\Sigma$.
 The $\Sigma$ is defined in the center-of-mass system as
\begin{equation}
 \label{eq:sigma}
 \frac{d\sigma}{d\Omega} = \frac{d\sigma_{0}}{d\Omega}(1 - P_{\gamma}\Sigma\cos{(2\Phi)})
\end{equation}
 where $\frac{d\sigma_{0}}{d\Omega}$ is the \textit{unpolarized} differential cross section,
 $P_{\gamma}$ is the degree of linear polarization of the photon beam, and
 $\Phi$ is the azimuthal angle between the linear polarization direction of the photon beam and the reaction plane of the $\eta$ photoproduction.
 $P_{\gamma}$ is calculated as a function of the photon beam energy by the formula based on the quantum electrodynamics \cite{BCS}.
 The photon beam asymmetry $\Sigma$ was determined by a fit to the yield distribution depending on the azimuthal angle $\Phi$:
\begin{equation}
 \label{eq:sigma_fit}
 f(\Phi) = A(1 + B\cos{(2\Phi)})
\end{equation}
 The fitting parameter \textit{B} in Eq. (\ref{eq:sigma_fit}) means the product of the photon beam polarization $P_{\gamma}$ and the photon beam asymmetry $\Sigma$.
 We used horizontally and vertically polarized photon beams alternately to reduce the systematic uncertainty arising from incomplete detector symmetry.
 The angles of these polarization vectors were estimated to be $-2.1$ and $82.6$ degrees from the horizontal plane in the laboratory frame, respectively.
 The degree of laser polarization was typically $98$\%.
 $P_{\gamma}$ was in the range of $42$--$91$\%, where the highest polarization was obtained at the Compton edge.
 The photon beam asymmetry was measured in 10 energy bins at $W$ = $1.82$--$2.32$ GeV and 8 polar angle bins at $\cos{\theta^{\eta}_{\mathrm{c.m.}}}$ = $-1.0$--$0.6$.
 At each kinematic bin, the sample was divided into 8 azimuthal-angle bins relative to the linear polarization vector of the photon beam.
\par

\begin{table}[!bp]
 \footnotesize
 \centering
 \caption{Systematic uncertainties of the photon beam asymmetry measurement}
 \begin{tabular}{l@{\hspace{10truemm}}l}\hline \hline
  Source of systematic uncertainties & Typical value \\ \hline
  Difference of two polarization data & $0.003$--$0.05$ \\
  Another binning of azimuthal angle & $0.004$--$0.05$ \\
  Ambiguity of polarization vector direction & $0.001$--$0.008$ \\
  Uncertainty of laser polarization degree & $0.04$\% of $|\Sigma|$\\ \hline \hline
 \end{tabular}
 \label{table:ba_sys}
\end{table}

 Systematic uncertainties for the measurement of the photon beam asymmetries are listed in Table \ref{table:ba_sys}.
 The listed numbers represent possible deviations in the $\Sigma$ values,
 and the estimated deviations are distributed in the indicated range depending on the kinematic bin.
 For the measurement of the following uncertainties, we combined neighboring kinematic bins to reduce the influence of statistical uncertainty.
 Firstly, the difference of the photon beam asymmetries in the horizontal and vertical polarization data
 was examined to conservatively treat it as a possible systematic uncertainty.
 Secondly, we considered an uncertainty due to different binning methods for azimuthal angles.
 This uncertainty was estimated by shifting a half bin at the azimuthal binning.
 Finally, the ambiguities of polarization vector direction and laser polarization degree were taken into account to estimate their influence on the photon beam asymmetries.
 The total systematic uncertainties were evaluated to be $0.008$--$0.09$ by taking a root of the quadratic sum of the above uncertainties.

\section{Results}
\subsection{Differential cross section}
\begin{figure*}[!tp]
 \centering
 \includegraphics[width=172truemm]{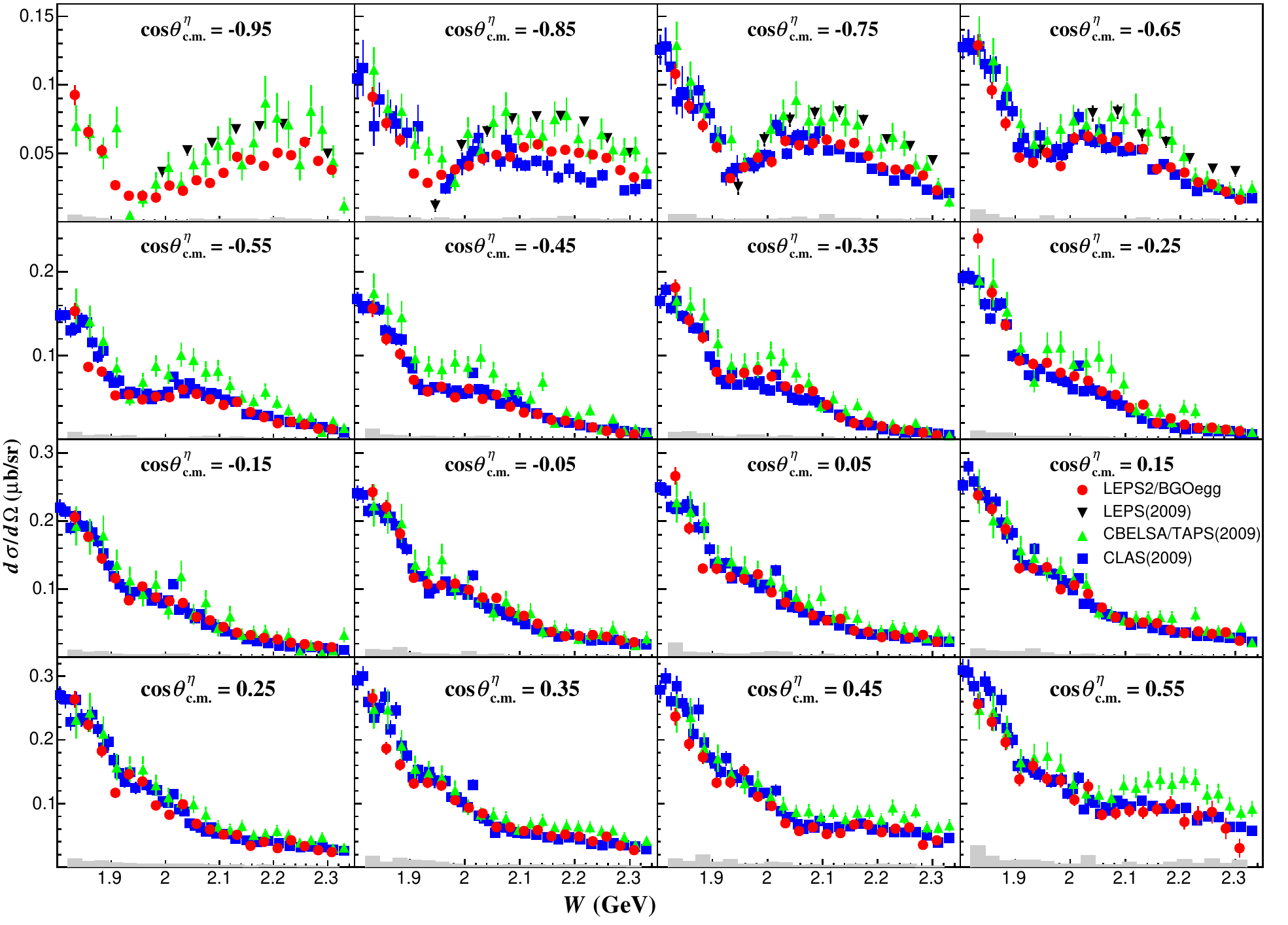}
 \caption{
 Differential cross sections $d\sigma / d\Omega$ as a function of $W$ for the reaction $\gamma p \rightarrow \eta p$.
 The individual panels correspond to different bins of the $\eta$ emission angle in the center-of-mass system.
 The present results are shown by red closed circles with statistical uncertainties.
 The green closed triangles, blue squares, and black inverted triangles come from other experimental results by
 the CBELSA/TAPS \cite{cbelsa2009}, CLAS \cite{clas2009}, and LEPS \cite{leps2009} collaborations, respectively.}
 \label{fig:dcs_energy_dep}
\end{figure*}

 We measured the differential cross sections for the reaction $\gamma p \rightarrow \eta p$ in $20$ total energy bins with 25-MeV steps
 and $16$ $\cos{\theta^{\eta}_{\mathrm{c.m}}}$ bins with $0.1$ steps.
 Figure \ref{fig:dcs_energy_dep} shows the energy dependence of differential cross sections
 measured by the present analysis and other experiments for the individual $\cos{\theta^{\eta}_{\mathrm{c.m}}}$ bins.
 The red solid circles are the results of the present analysis with statistical uncertainties, and the gray histograms are the associated systematic uncertainties.
 The black triangles, green triangles, and blue squares show the results of the LEPS \cite{leps2009}, CBELSA/TAPS \cite{cbelsa2009}, and CLAS \cite{clas2009} experiments, respectively.
 The LEPS and CBELSA/TAPS results have been obtained for the photon beam energy bins of each 100 and 50 MeV, respectively.
 The CLAS results have been obtained for the total energy bins of each $10$ MeV at $W$ = $1.68$--$2.1$ GeV and each $5$ MeV at $W$ = $2.1$--$2.36$ GeV.
 All of those results are consistently binned with $0.1$ steps in $\cos{\theta^{\eta}_{\mathrm{c.m}}}$.
\par

 The present analysis for the BGOegg experiment has achieved the precise and wide angular measurement
 by detecting all the final states including a proton and an $\eta$ meson, which decays into $\gamma\gamma$.
 Although the proton momentum was treated as an unmeasured variable in the kinematic fit to obtain the results in Fig.~\ref{fig:dcs_energy_dep},
 the validity of the procedure was confirmed by the independent cross-section measurement using the RPC, as discussed later.
 In contrast, the LEPS experiment used a missing mass technique by measuring only the proton momentum and emission angle
 in the limited acceptance $\cos{\theta^{\eta}_{\mathrm{c.m}}}<-0.6$, as mentioned in Sec.~I.
 The CLAS experimental setup was optimized for the detection of charged particles,
 so the identification of the $\eta$ meson was done using the $\eta\to\pi^{+}\pi^{-}\pi^0$ decay mode,
 where the $\pi^0$ was treated as a missing particle in the kinematic fit.
 The CBELSA/TAPS experiment analyzed $\eta$ decays into the two modes of
 $\eta \rightarrow \gamma\gamma$ and $\eta \rightarrow 3\pi^0 \rightarrow 6\gamma$ by using large acceptance calorimeters,
 but the statistics are limited compared with other experiments.

\par

 The present results by the BGOegg experiment (red solid circles) generally show a declining trend of differential cross sections
 as the energy increases in the region of $\cos{\theta^{\eta}_{\mathrm{c.m.}}}>0$.
 A bump structure appears in the region of $\cos{\theta^{\eta}_{\mathrm{c.m.}}}<0$, and its strength becomes larger as the $\eta$ emission angles get more backward.
 This bump position is around $W$ = $1.97$ GeV at $-0.1<\cos{\theta^{\eta}_{\mathrm{c.m.}}}<0$ and slightly shifts to $W$ = $2.02$ GeV at $-0.7<\cos{\theta^{\eta}_{\mathrm{c.m.}}}<-0.6$.
 The peak position changes more rapidly at the most backward angles, and is around $W$ = $2.25$ GeV at $-1<\cos{\theta^{\eta}_{\mathrm{c.m.}}}<-0.9$.
\par

\begin{figure}[!bp]
 \centering
 \includegraphics[width=86truemm]{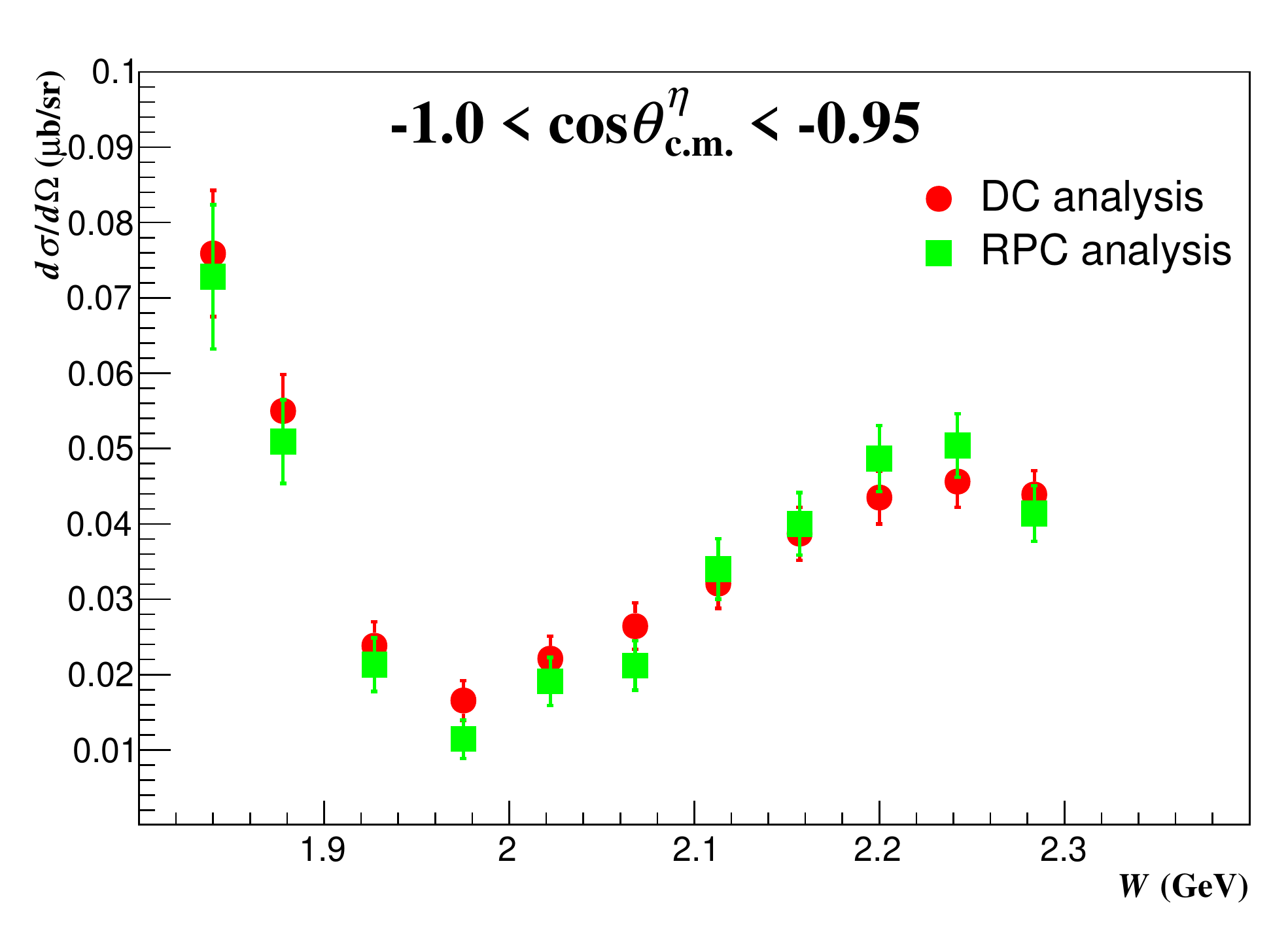}
 \caption{
 Comparison of the differential cross sections with and without the use of RPC for the extremely backward $\eta$ angles $-1.0 < \cos{\theta^{\eta}_{\mathrm{c.m.}}} < -0.95$.
 The red closed circles and green squares are the results by using the DC only and both the RPC and DC, respectively.}
 \label{fig:dcs_dc_rpc}
\end{figure}

\begin{figure*}[!tp]
 \centering
 \includegraphics[width=172truemm]{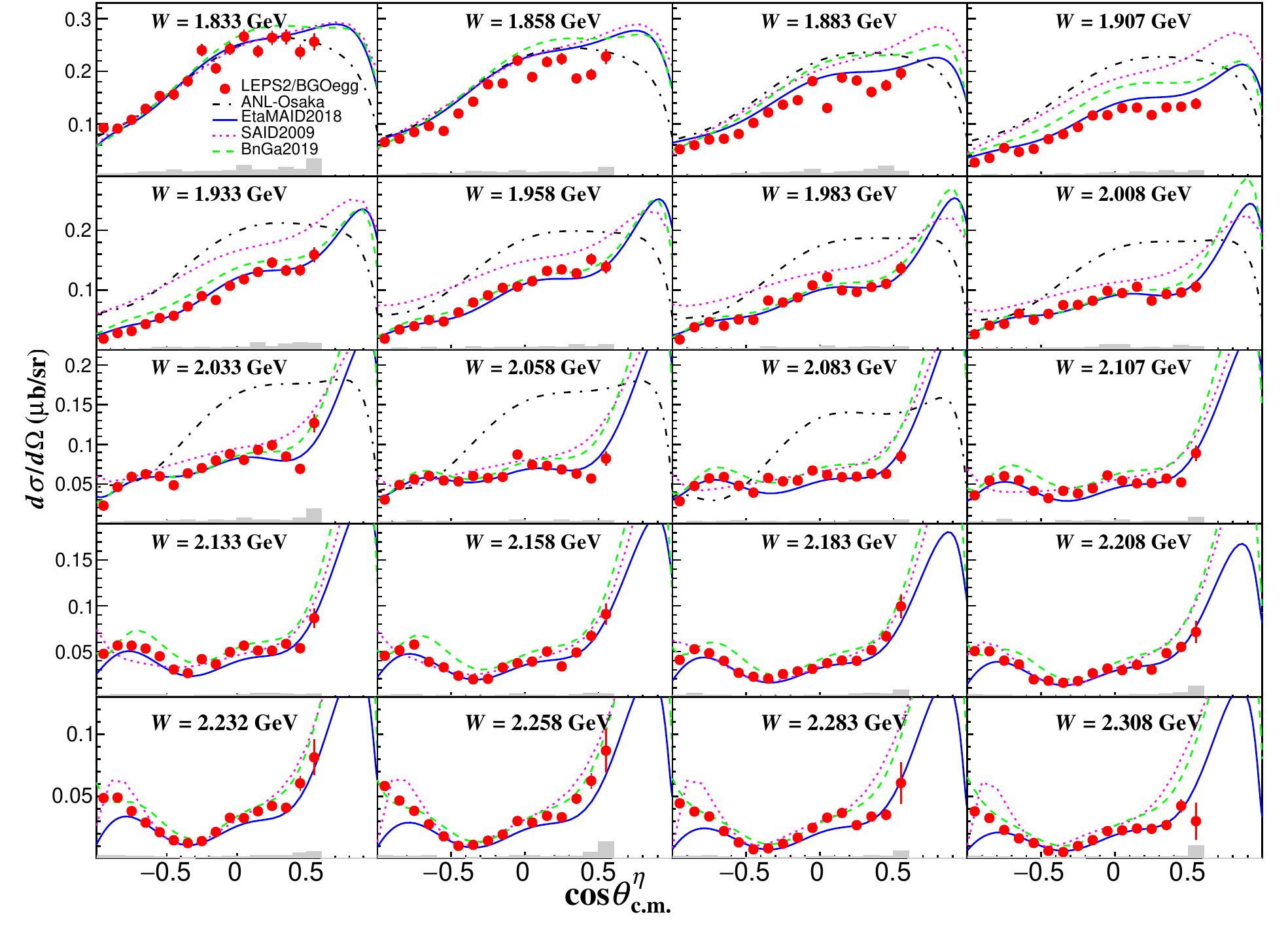}
 \caption{
 Differential cross sections $d\sigma / d\Omega$ as a function of $\cos{\theta^{\eta}_{\mathrm{c.m.}}}$ for the reaction $\gamma p \rightarrow \eta p$.
 The present results are shown by red closed circles with statistical uncertainties.
 Estimated systematic uncertainties are indicated by the gray histograms.
 The blue solid, magenta dotted, green dashed, and black dash-dotted curves show the PWA model calculations by EtaMAID2018 \cite{maid2018}, SAID2009 \cite{said_web}, Bonn-Gatchina2019 \cite{bnga2019}, and ANL-Osaka2016 \cite{anl_osaka}, respectively.
 }
 \label{fig:dcs_angle_dep}
\end{figure*}

 As a whole, the BGOegg results well agree with the CLAS data for $\cos{\theta^{\eta}_{\mathrm{c.m.}}} > -0.8$.
 The CBELSA/TAPS and LEPS data give larger cross sections compared to the BGOegg and CLAS results.
 The bump structure is seen in all the experiments while its shape and strength are different among them.
 For instance, there is a discrepancy in the bump structure shape between the BGOegg and CLAS data at $-0.9<\cos{\theta^{\eta}_{\mathrm{c.m.}}}<-0.8$,
 which however corresponds to the acceptance boundary of the CLAS measurement.
 The BGOegg result agrees with the LEPS data in terms of the peak position at the $\eta$ angles $-0.9<\cos{\theta^{\eta}_{\mathrm{c.m.}}}<-0.8$,
 but the amplitude of the bump in the LEPS measurement is inconsistently higher.
 At extremely backward angles $-1.0<\cos{\theta^{\eta}_{\mathrm{c.m.}}}<-0.9$, both the peak position and strength of the bump are inconsistent between the BGOegg and LEPS results.
 The CBELSA/TAPS data at the corresponding $\eta$ angles also show the bump structure but have large uncertainties,
 which make it difficult to examine the strength and shape of the observed structure in detail.
\par

 Because there are discrepancies in the differential cross section results among the different experiments at the extremely backward $\eta$ angles,
 it is important to confirm our measurement in a more precise manner.
 In order to provide validity to our result, an independent analysis of the same data set was additionally performed by detecting a proton at the RPC.
 The RPC can measure the momentum of a forward proton via its time-of-flight at the extremely backward $\eta$ angles ($-1.0 < \cos{\theta^{\eta}_{\mathrm{c.m.}}} < -0.95$).
 Thus, it allows the complete 4-momentum conservation constraints to be used in the kinematic fit without unmeasured variables.
 Figure \ref{fig:dcs_dc_rpc} shows the comparison of differential cross sections obtained by using the RPC (green squares)
 and the same procedure as done for Fig. \ref{fig:dcs_energy_dep} only with the DC (red circles) in the overlapping acceptance region.
 We confirmed a very good agreement between the two analyses both in the energy dependence and overall magnitude of the differential cross sections.
\par

\begin{figure*}[!tp]
 \centering
 \includegraphics[width=172truemm]{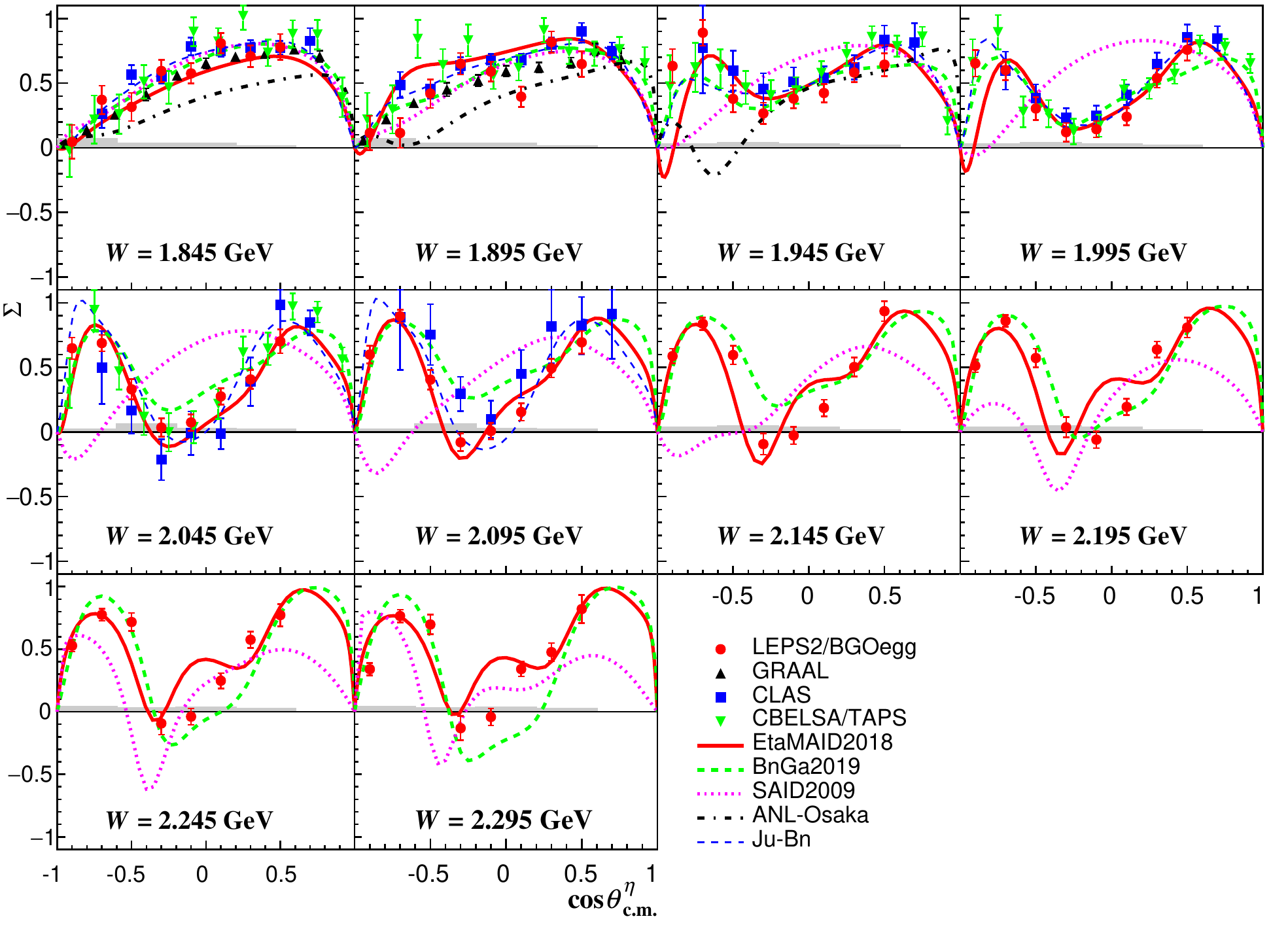}
 \caption{Photon beam asymmetries $\Sigma$ as a function of $\cos{\theta^{\eta}_{\mathrm{c.m.}}}$ for the reaction $\gamma p \rightarrow \eta p$.
          The present results are shown by the red closed circles with statistical uncertainties and
          the associated systematic uncertainties are shown by the gray histograms.
          The black closed triangles, blue squares, and green inverted triangles come from other experimental results
          by GRAAL \cite{graal}, CLAS \cite{clas2017}, and CBELSA/TAPS \cite{cbelsa2020} collaborations, respectively.
          The blue solid, green dashed, magenta dotted, black dash-dotted, and blue long-dashed curves represent the PWA results
          by the EtaMAID2018 \cite{maid2018}, Bonn-Gatchina2019 \cite{bnga2019}, SAID2009 \cite{said_web},
          ANL-Osaka \cite{anl_osaka}, and J\"{u}lich-Bonn \cite{jubn} models, respectively.}
 \label{fig:ba_angle_dep}
\end{figure*}

 The angular distributions of differential cross sections for different energy bins are shown in Fig.~\ref{fig:dcs_angle_dep}.
 The present results (red points) show a backward rise in the higher energy region.
 We compared our results with the PWA calculations by EtaMAID2018 \cite{maid2018,maid_web} (blue solid lines),
 SAID2009 \cite{said_web} (magenta dotted lines), Bonn-Gatchina2019 \cite{bnga2019,bnga_web} (green dashed lines), and ANL-Osaka2016 \cite{anl_osaka, anl_osaka_2} (black dotted-dashed lines).
 The measured data are consistent with the EtaMAID2018 prediction in the total energy region below $2.2$ GeV,
 while the EtaMAID2018 can not reproduce our data at the most backward angles in the case of total energies above $2.2$ GeV.
 The SAID calculations are overestimated compared to our differential cross sections in the region of $W$ = $1.9$--$2.0$ GeV,
 and this disagreement disappears at the higher energies.
 However, the peaking structure of the SAID calculation at $\cos{\theta^{\eta}_{\mathrm{c.m.}}}<-0.6$ is not observed in our data.
 The Bonn-Gatchina2019 calculations are more or less in agreement with our data,
 reproducing the enhancement of differential cross sections at the backward angles.
 This is because their PWA fit utilizes all the data that are recently available except for the present results.
 The validity of the Bonn-Gatchina2019 model can be examined by our photon beam asymmetry result, which is the first measurement at higher energies as described in the next subsection.

\subsection{Photon beam asymmetry}
 We measured the photon beam asymmetries for the reaction $\gamma p \rightarrow \eta p$ in 10 total energy bins with 50-MeV steps
 and 8 $\cos{\theta^{\eta}_{\mathrm{c.m}}}$ bins with 0.2 steps.
 The red closed circles in Fig.~\ref{fig:ba_angle_dep} show the measured photon beam asymmetries $\Sigma$ with statistical uncertainties as a function of $\cos{\theta^{\eta}_{\mathrm{c.m.}}}$,
 and the gray histograms are the associated systematic uncertainties.
 Each data point is plotted at the mean $\cos{\theta^{\eta}_{\mathrm{c.m.}}}$ value of entries in the corresponding angular bin.
 In Fig.~\ref{fig:ba_angle_dep}, other experimental results from the GRAAL \cite{graal}, CLAS \cite{clas2017}, and CBELSA/TAPS \cite{cbelsa2020} collaborations
 are also compared with the present results by the BGOegg experiment.
 Here all the overlaid results have used different energy-binning methods.
 The GRAAL results are divided into 15 photon beam energy bins in the $E_{\gamma}$ range of $0.7$--$1.5$ GeV,
 while the CBELSA/TAPS results have been obtained for the photon beam energy bins of each $60$ MeV in the range of $E_{\gamma}$ = $1.13$--$1.79$ GeV.
 The CLAS experiment has adopted the photon beam energy bins of $27$ and $40$ MeV at $1.071 < E_{\gamma} < 1.689$ and $1.689 < E_{\gamma} < 1.876$ GeV, respectively.
 In Fig.~\ref{fig:ba_angle_dep}, these results are plotted at the energies that are closest to the energy bins of the individual analyses.
\par

 The present results by the BGOegg experiment statistically agree with the other experimental results in the overlapped energy region below $W$ = $2.1$ GeV.
 The photon beam asymmetry have a dip structure around $\cos{\theta^{\eta}_{\mathrm{c.m.}}} = -0.2$ at $W > 1.9$ GeV.
 It has been suggested that this behavior is influenced by the helicity couplings for $N (1720) 3/2^{+}$ and $N (1900) 3/2^{+}$ \cite{clas2017}.
 The precise $\Sigma$ values in a wide angular range were obtained for the first time above the total energy of about $2.1$ GeV.
 The dip structure remains at higher energies.
\par

 The overlaid curves in Fig.~\ref{fig:ba_angle_dep} show the existing PWA results calculated by
 the EtaMAID2018 \cite{maid2018}, Bonn-Gatchina2019 \cite{bnga2019}, SAID2009 \cite{said_web}, ANL-Osaka \cite{anl_osaka}, and J\"{u}lich-Bonn \cite{jubn} models.
 The ANL-Osaka and J\"{u}lich-Bonn results are limited to the total energy ranges below $1.95$ and $2.1$ GeV, respectively.
 The ANL-Osaka does not reproduce the experimental data in all energy regions
 because it does not included heavy-meson contributions such as an $\omega$ meson in the coupled-channel calculation.
 The SAID does not reproduce the dip structure above $W$ = $1.95$ GeV.
 The EtaMAID2018, Bonn-Gatchina2019, and J\"{u}lich-Bonn models agree with the present results in the total energy region below $2.0$ GeV except for the extremely backward region.
 In the region above $2.0$ GeV, no PWA results reproduce the BGOegg results.

\section{Discussion}
\subsection{Differential cross section enhancement at $W$ = $2.0$--$2.3$ GeV}
 The angular dependence of the differential cross section above $W$ = $2.1$ GeV in Fig.~\ref{fig:dcs_angle_dep} shows an enhancement at $\cos{\theta^{\eta}_{\mathrm{c.m.}}} < -0.4$,
 where one can expect the possible contributions from a \textit{u}-channel exchange or high-spin \textit{s}-channel resonances.
 Regge theory \cite{Regge,JKStorrow} allows us to assume a simple description of the smooth energy dependence for the \textit{u}-channel cross section in the form of $s^{2\alpha(u)-2}$,
 where $s$ and $\alpha(u)$ denote the center-of-mass energy and a Regge trajectory function, respectively.
 Therefore the bump-like energy dependence in a narrow range of $2.0 < \textrm{W} < 2.4$ GeV, as shown in Fig.~\ref{fig:dcs_energy_dep},
 cannot be explained only by the \textit{u}-channel contribution.
 The value of $(2\alpha(u)-2)$ is also expected to be negative in a small $\left| u \right|$ region as shown in Fig.~33 of the Ref.~\cite{Laget}.
 In addition, the EtaMAID2018 calculation describes the non-resonant background as \textit{s}- and \textit{u}-channel Born terms and \textit{t}-channel vector meson exchanges.
 This calculation hints that the amplitude of the \textit{u}-channel contribution is rather small \cite{maid2018}.
 Our data suggest that the steep backward rise of differential cross sections is likely related to the decay of high-spin \textit{s}-channel resonances.
 In the photon-proton reaction, the helicity of the initial state is limited to $\left| h \right| \leq 3/2$.
 Therefore, if an intermediate resonance has a high spin ($J \geq 5/2$),
 it can emit an $\eta$ meson to the backward or forward polar angles in two-body decays,
 as understood by the discussion of helicity amplitudes with Wigner d-matrices \cite{helicity_amplitude}.
 The differential cross sections in the backward $\eta$ angles are more sensitive to the high-spin \textit{s}-channel resonances
 because of the suppression of \textit{t}-channel meson exchanges.
\par

\begin{figure}[!tbp]
 \centering
 \includegraphics[width=60truemm]{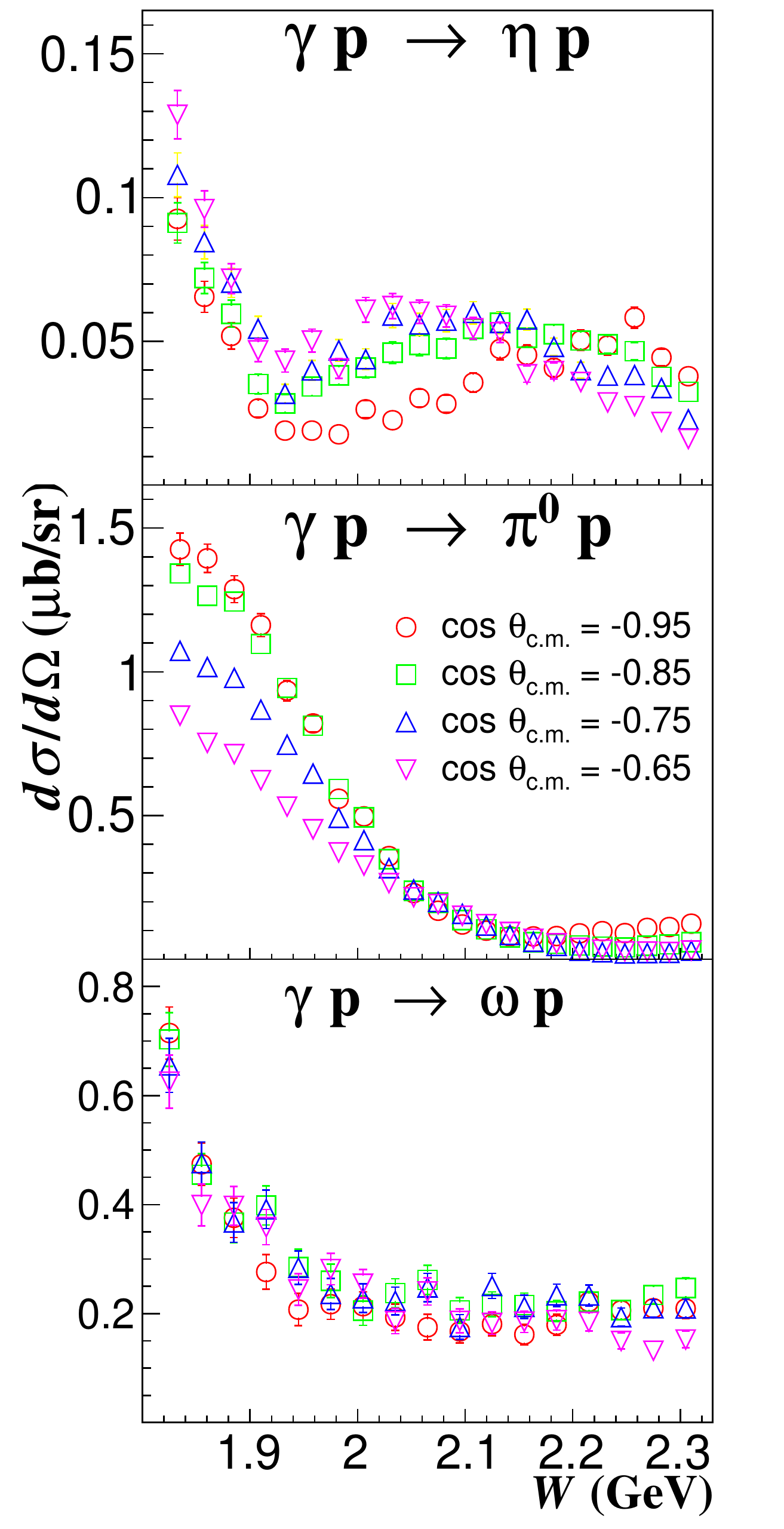}
 \caption{The differential cross sections of the $\eta$, $\pi^0$ and $\omega$ photoproduction processes as a function of the total energy W
          at the angles $\cos{\theta_{\mathrm{c.m.}}} = -0.95$ (red circles), $-0.85$ (green squares),
          $-0.75$ (blue triangles), $-0.65$ (magenta inverted triangles).}
 \label{fig:dcs_comp}
\end{figure}

 Figure \ref{fig:dcs_comp} shows the energy dependence of differential cross sections at $\cos{\theta_{\mathrm{c.m.}}}=-0.95$, $-0.85$, $-0.75$, and $-0.65$
 for the $\eta$, $\pi^0$ and $\omega$ photoproduction processes in the BGOegg experiment.
 The differential cross sections of the $\pi^0$ and $\omega$ photoproduction were obtained using the same data set
 as the present analysis and reported in Refs.~\cite{eggpi0} and \cite{eggomega}, respectively.
 The differential cross section distributions of the $\pi^0$ photoproduction show declining behaviors from $1.8$ to $2.1$ GeV.
 At the angle $\cos{\theta^{\pi^0}_{\mathrm{c.m.}}}=-0.95$, a small enhancement above $2.1$ GeV is seen, but there is no visible bump structure.
 The differential cross sections of the $\omega$ photoproduction also show no structures above $1.9$ GeV.
 In contrast, only the $\eta$ photoproduction shows a clear bump structure at the total energies above $2.0$ GeV.
 In the flavor SU(3) quark models, the $\eta$ meson contains $s\bar{s}$ quark pair in its composition
 while the $\pi^0$ and $\omega$ mesons have flavor configurations only with $u\bar{u}$ and $d\bar{d}$ quarks.
 Therefore, the observed bump structure in the differential cross sections of $\eta$ photoproduction is likely associated with the nucleon resonances
 that have a large $s\bar{s}$ component and strongly couple to the $\eta N$ channel.
\par

 In the total energy dependences of differential cross sections, the position of the bump structure shifts
 from $2.02$ GeV at $\cos{\theta^{\eta}_{\mathrm{c.m.}}} = -0.65$ to $2.25$ GeV at $\cos{\theta^{\eta}_{\mathrm{c.m.}}} = -0.95$, as mentioned in Sec.~V-A.
 This may be caused by the presence of multiple resonances with the isospin 1/2.
 In the mass range of $2.1$--$2.3$ GeV, several resonances with three or four stars are currently known based on the $\pi N$-decay channel
 (e.g. N(2100)1/$2^+$, N(2120)3/$2^-$, N(2190)7/$2^-$, N(2220)9/$2^+$, N(2250)9/$2^-$) \cite{Zyla:2020zbs}.
 However, the information about $\eta N$-decay channel of nucleon resonances is limited.
 Hence, the new BGOegg data of differential cross sections will provide additional constraints for the resonance search,
 particularly with high precisions at backward $\eta$ angles.
\par

\subsection{Comparison with the existing PWA results}
 Our experimental differential cross sections and photon beam asymmetries more or less agree with the existing PWA results at lower energies.
 In contrast, the PWA results at higher energies show clear differences from our experimental data, as described in Sec.~V.
 The discrepancies in the photon beam asymmetries are particularly large.
 In addition, the PWA results are inconsistent with each other at the higher energies.
\par

\begin{figure*}[!tp]
 \centering
 \includegraphics[width=150truemm]{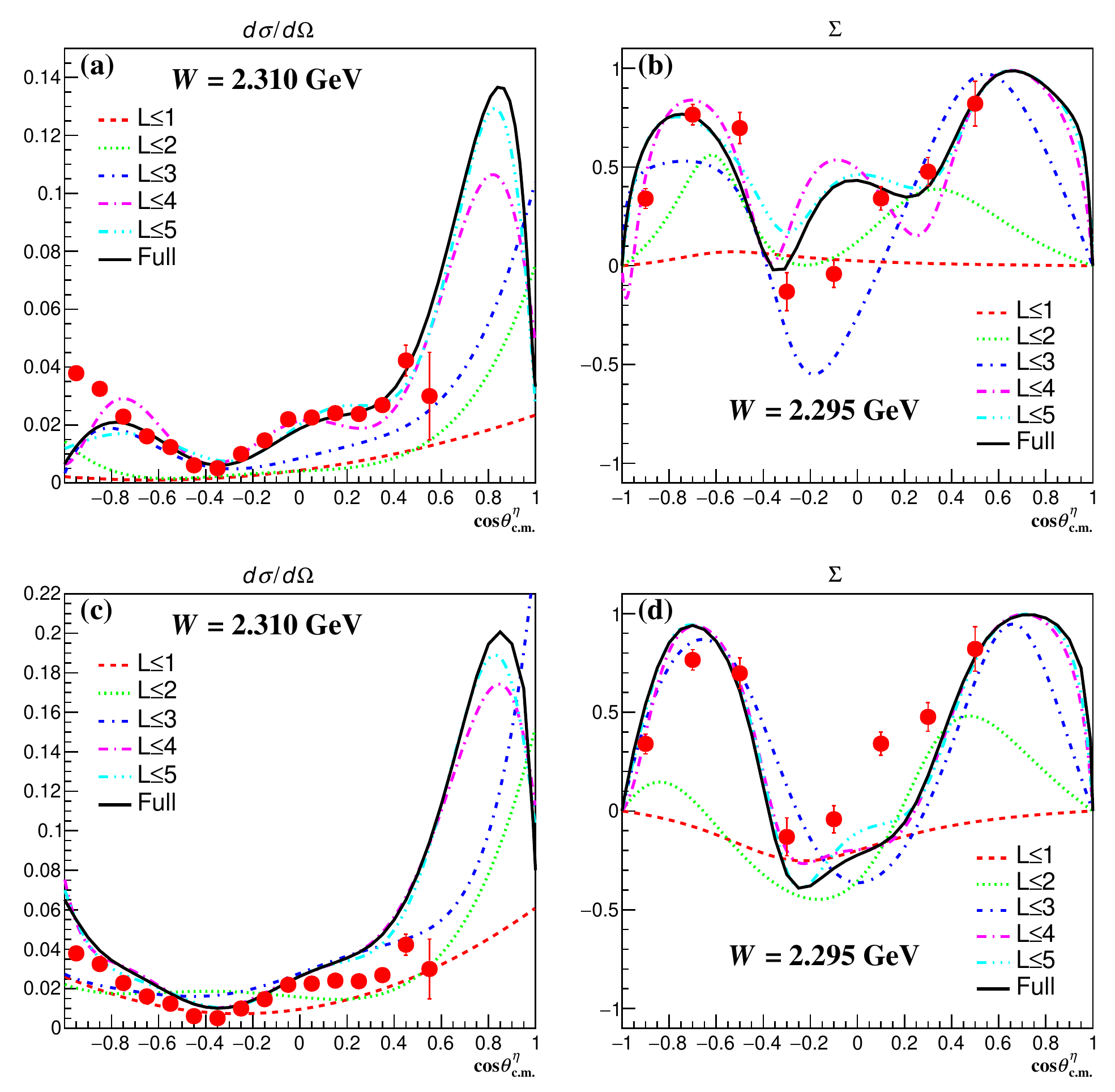}
 \caption{The existing PWA results calculated by EtaMAID2018 ((a) and (b)) and Bonn-Gatchina2019 ((c) and (d))
          with the orbital angular momenta \textit{L} up to 1 (red dashed lines), 2 (green dotted lines), 3 (blue dotted-dashed lines),
          4 (magenta long dashed-dotted lines), and 5 (cyan dotted lines).
          The full PWA calculation with all the orbital angular momenta is shown by black solid lines.
          The BGOegg results are plotted as red circles.}
 \label{fig:multi}
\end{figure*}
\par

 Figure \ref{fig:multi} shows the comparison of our experimental data and
 the existing PWA results calculated by the EtaMAID2018 ((a) and (b)) and Bonn-Gatchina2019 ((c) and (d)) models around $W$ = $2.3$ GeV,
 which corresponds to the highest energy bin in our measurement.
 Differential cross sections and photon beam asymmetries are plotted in the left ((a) and (c)) and right ((b) and (d)) sides, respectively.
 We compared the model calculations in the various ranges of orbital angular momenta \textit{L}.
 In the PWA of pseudoscalar-meson photoproduction, the Chew-Goldberger-Low-Nambu (CGLN) amplitudes \cite{cgln} are conventionally used.
 These amplitudes are simply represented by using electromagnetic multipoles with Legendre polynomials.
 These electromagnetic multipoles include the information about the partial wave of a meson-nucleon system.
 The electromagnetic multipole amplitudes of EtaMAID2018 and Bonn-Gatchina2019 were obtained from Refs.~\cite{maid_web} and \cite{bnga_web}, respectively.

 The measured differential cross sections are well reproduced by the EtaMAID2018 full calculation except in the most backward $\eta$-angle region.
 The observed backward rise of differential cross sections does not exist in EtaMAID2018 results.
 The calculated photon beam asymmetry has a small bump structure at $\cos{\theta^{\eta}_{\mathrm{c.m.}}} \sim 0$.
 This structure is not seen in the experimental results.
 On the other hand, the Bonn-Gatchina2019 results reproduce the backward shape of differential cross sections
 but its strength is overestimated compared to our results.
 The calculated photon beam asymmetry has no small bump structure like the EtaMAID2018 calculation
 but it is underestimated compared to our results around $\cos{\theta^{\eta}_{\mathrm{c.m.}}} \sim 0$.
\par

 Our differential cross sections show a sharp backward rise at higher energies.
 By comparing this behavior with the two PWA results in Fig.~\ref{fig:multi},
 it is recognized that the determination of multipole contributions at \textit{L} $\leq$ $3$ is still ambiguous in the existing PWAs
 and important to reproduce the data.
 The different determinations of multipoles between PWA calculations in the lower \textit{L} region also make a large difference in the angular dependence of calculated photon beam asymmetries.
 This indicates that the current understanding of resonance and born-term contributions is not enough
 even for lower \textit{L}'s at high energies in both the PWA calculations.
 In addition, higher \textit{L} contributions are important to accurately reproduce the measured photon beam asymmetries.
\par

 The J\"{u}lich-Bonn model curves have been determined by a fit to the CLAS results
 of differential cross sections and photon beam asymmetries in the $\eta$ photoproduction \cite{clas2017}.
 In the CLAS measurement, the photon beam asymmetries were obtained at $1.70 < \textrm{W} < 2.10$ GeV and $-0.8 < \cos{\theta^{\eta}_{\mathrm{c.m.}}} < 0.8$.
 Before the fit was made, the $N(1900)3/2^{+}$ was found to be important
 in the analyses of $K\Lambda$ and $K\Sigma$ photoproduction by the Bonn-Gatchina group \cite{kphoto}.
 In order to confirm this resonance contribution, the CLAS collaboration fitted two sets of possible solutions
 with and without a contribution from the $N(1900)3/2^{+}$ resonance by using the J\"{u}lich-Bonn model.
 They discussed the weakness of the $N(1900)3/2^{+}$ contribution in the $\eta$ photoproduction but was not able to clearly determine its strength
 because the difference of two fits should appear in the photon beam asymmetries at extremely backward $\eta$ angles,
 which were out of the CLAS measurement range.
 In contrast, we have obtained experimental results including the photon beam asymmetries at the most backward angles.
 Note that the BGOegg and CLAS results are consistent with each other in the overlapped angular region.
 A refit of the J\"{u}lich-Bonn model to our new data will provide more accurate information about the strength of the \textit{N}(1900) contribution.

\section{Summary}
 We measured differential cross sections and photon beam asymmetries for the reaction $\gamma p \to \eta p$ by detecting $\eta \to \gamma\gamma$ decay mode.
 The photon beam is produced by backward Compton scattering in the energy range of $1.3$--$2.4$ GeV at the SPring-8 LEPS2 beamline.
 This photon beam is linearly polarized and the degree of polarization is more than $90$\% at the Compton edge.
 The two $\gamma$'s in the final state were measured using the BGOegg calorimeter, which has large acceptance and the world's best energy resolutions.
 The direction of a proton in the final state was measured using the BGOegg calorimeter or the DC.
 To select a signal sample, we applied a kinematic fit using the 4-momenta of two $\gamma$'s, the direction of a final-state proton,
 the photon beam energy measured using the tagger, and a vertex position.
 The background estimation was done by the template fitting.
\par

 The differential cross sections and photon beam asymmetries were derived in the kinematic bins of total energies and $\eta$ polar angles
 covering $1.82$--$2.32$ GeV and $-1.0 \leq \cos{\theta^{\eta}_{\mathrm{c.m.}}} \leq 0.6$, respectively.
 The validity of our cross section measurement was confirmed by an independent analysis using the RPC, which additionally measured the momentum of a forward proton.
 A bump structure appears at $W$ = $2.02$--$2.25$ GeV in the case of $\cos{\theta^{\eta}_{\mathrm{c.m.}}} < 0$,
 and its strength becomes larger as the $\eta$ emission angles get more backward.
 The bump structure is seen in the LEPS, CBELSA/TAPS, and CLAS experiments
 but their shapes and strengths are different among these experiments.
 Our new measurement of differential cross sections provides high-precision and reliable data in the backward angular region.
 The bump-like enhancement indicates the contribution of high-spin nucleon resonances that contain a large $s\bar{s}$ component.
 The peak position of the bump structure moves depending on the $\eta$ emission angle, suggesting the contribution of multiple resonances.
 For the first time, we measured the photon beam asymmetry of the $\eta$ photoproduction above $W$ = $2.1$ GeV.
 No PWA calculations reproduce our results in the higher energy region.
 The multipole amplitudes with even low orbital angular momenta are different between the existing PWA models.
 Our new results will provide  additional constraints for the understanding of baryon resonances via PWAs.

\section*{Acknowledgments}
 The experiment was performed at the BL31LEP of SPring-8 with the approval of the Japan Synchrotron Radiation Institute (JASRI) as a contract beamline (Proposal No. BL31LEP/6101).
 The authors are grateful to the staff at Spring-8 for supporting the commissioning of the LEPS2 beamline and providing the excellent experimental conditions during the data collection.
 We thank T. Sato, A. Sarantsev, V. Nikonov, K. Nikonov, and D. R\"{o}nchen for discussions on the partial wave analyses.
 We appreciate A. Hosaka for discussions about the amplitude calculations.
 This research was supported in part by Ministry of Education, Culture, Sports, Science and Technology of Japan (MEXT),
 Scientific Research on Innovative Areas Grant No.~JP21105003 and No.~JP24105711,
 Japan Society for the Promotion of Science (JSPS) Grant-in-Aid for Specially Promoted Research Grant No.~JP19002003,
 Grant-in-Aid for Scientific Research (A)Grant No.~JP24244022,
 Grant-in-Aid for Young Scientists (A) Grant No.~JP16H06007,
 Grants-in-Aid for JSPS Fellows No.~JP24608,
 and the Ministry of Science and Technology of Taiwan.


\begin{thebibliography}{00}
 \bibitem{Zyla:2020zbs}
 P.~A.~Zyla \textit{et al.}, PTEP \textbf{2020}, 083C01 (2020)
 \bibitem{cqm}
  S. Capstick and W. Roberts, Prog. Part. Nucl. Phys. \textbf{45}, S241 (2000).
 \bibitem{RoperN1440}
 L.~D.~Roper, Phys. Rev. Lett. \textbf{12}, 340-342 (1964)
 \bibitem{leps2009}
 M. Sumihama, \textit{et al}., Phys. Rev. C \textbf{80}, 052201 (2009).
 \bibitem{clas2009}
 M. Williams, \textit{et al}., Phys. Rev. C \textbf{80}, 045213 (2009).
 \bibitem{cbelsa2009}
 V. Crede, \textit{et al}., Phys. Rev. C \textbf{80}, 055202 (2009).
 \bibitem{clas2017}
 P. Collins, \textit{et al}., Phys. Lett, B \textbf{771}, 213 (2017).
 \bibitem{cbelsa2020}
 F. Afzal, \textit{et al}., Phys. Rev. Lett. \textbf{125}, 152002 (2020).
 \bibitem{leps2}
 N. Muramatsu, \textit{et al}., arXiv:2112.07832.
 \bibitem{beamline}
 N. Muramatsu, \textit{et al}., Nucl. Instr. Meth. A \textbf{737}, 184 (2014).
 \bibitem{bgoegg_nim}
 T. Ishikawa, \textit{et al}., Nucl. Instr. Meth. A \textbf{837}, 109 (2016).
 \bibitem{rpc_jinst}
 N. Tomida, \textit{et al}., JINST \textbf{9}, C10008 (2014).
 \bibitem{rpc_jinst2}
 N. Tomida, \textit{et al}., JINST \textbf{11}, C11037 (2016).
 \bibitem{eggpi0}
 N. Muramatsu, \textit{et al}., Phys. Rev. C \textbf{100}, 055202 (2019).
 \bibitem{SP8filling}
 http://www.spring-8.or.jp/ja/users\/operation\_status/
 schedule/bunch\_mode
 \bibitem{geant4}
 S. Agostinelli, \textit{et al}., Nucl. Instr. Meth. A \textbf{506}, 250 (2003);
 J. Allison, \textit{et al}., IEEE Trans. Nucl. 816 Sci. \textbf{53}, 270 (2006).
 \bibitem{BCS}
A. D’Angelo, O. Bartalini, V. Bellini, P. Levi Sandri, D. Moricciani, L. Nicoletti and A.
Zucchiatti, Nucl. Instr. Meth. A455 (2000) 1.
 \bibitem{maid2018}
 L. Tiator, \textit{et al}., Eur. Phys. J. A \textbf{54}, 210 (2018).
 \bibitem{maid_web}
 https://maid.kph.uni-mainz.de/eta2018/etamaid2018.html
 \bibitem{said_web}
 http://gwdac.phys.gwu.edu/
 \bibitem{bnga2019}
 CBELSA/TAPS collaboration (J. M\"{u}ller, \textit{et al}.), Phys. Lett. B \textbf{803}, 135323 (2020).
 \bibitem{bnga_web}
 https://pwa.hiskp.uni-bonn.de/
 \bibitem{anl_osaka}
 H. Kamano, S. X. Nakamura, T-S H. Lee, and T.Sato, Phys. Rev. C \textbf{94} 015201 (2016).
 \bibitem{anl_osaka_2}
 Private communication with T. Sato
 \bibitem{graal}
 O. Bartalini, \textit{et al}., Eur. Phys. J. A \textbf{33}, 169 (2007).
 \bibitem{jubn}
  D. R\"{o}nchen, \textit{et al}., Eur. Phys. J. A \textbf{51}, 70 (2015).
 \bibitem{Regge}
 P.D.B. Collins, \textit{An Introduction to Regge Theory and High Energy Physics}, Cambridge, Univ. Press. (1977).
 \bibitem{JKStorrow}
 J.K. Storrow, Rep. Prog. Phys. \textbf{50}, 1229, (1987).
 \bibitem{Laget}
 J.M. Laget, Prog. Part. Nucl. Phys. \textbf{111}, 103737 (2020).
 \bibitem{helicity_amplitude}
 C. G. Fasano, Frank Tabakin, and Bijan Saghai, Phys. Rev. C \textbf{46} 2430 (1992)
 \bibitem{eggomega}
 N. Muramatsu, \textit{et al}., Phys. Rev. C \textbf{102}, 025201 (2020).
 \bibitem{cgln}
 G.F. Chew, M.L. Goldberger, F.E. Low, and Y. Nambu, Phys. Rev. \textbf{106}, 1345 (1957).
 \bibitem{kphoto}
 A.V Anisovich, R. Beck, E. Klempt, V.A. Sarantsev, U. Thoma, Eur. Phys. J. A \textbf{48}, 88 (2012).
\end{thebibliography}
\end{document}